\documentclass[aps,prd,onecolumn,superscriptaddress,notitlepage,nofootinbib,longbibliography]{revtex4-2}

\usepackage{amsmath,amssymb,mathtools,physics,bm,hyperref}
\usepackage{microtype}
\usepackage{booktabs}
\usepackage{graphicx}
\usepackage{tikz}
\usepackage[compat=1.1.0]{tikz-feynman}
\usepackage{slashed}
\usepackage{framed}
\usepackage{titlesec}
\usepackage{cleveref}
\usepackage{tikz-feynhand}
\usepackage{subcaption}
\usetikzlibrary{decorations.markings}
\usetikzlibrary{arrows.meta, decorations.markings, calc}
\usepackage{tcolorbox}
\usepackage{amsthm}

\usepackage{gensymb}
\usepackage[utf8]{inputenc}
\usepackage{xcolor}
\usepackage{ulem}
\usepackage{float}
\usepackage{stackengine}

\providecommand{\VEV}[1]{\left\langle #1 \right\rangle}

\newcommand{\be}{\begin{equation}}
\newcommand{\ee}{\end{equation}}
\newcommand{\bea}{\begin{eqnarray}}
\newcommand{\eea}{\end{eqnarray}}

\hypersetup{
    colorlinks=true,
    linkcolor=blue,
    filecolor=magenta,
    urlcolor=cyan,
    citecolor=blue,
    pdfauthor={Mussawir Khan},
    pdftitle={Soft Breaking Naturalness in Majorana Fermion Loop induced Scalar and Majoron Amplitudes}
}

\begin{document}

\title{Soft Breaking Naturalness in Majorana Fermion Loop induced Scalar and Majoron Amplitudes}
\author{Mussawir Khan}
\affiliation{State Key Laboratory of Particle Astrophysics, Institute of High Energy Physics, Chinese Academy of Sciences, Beijing 100049, China}
\affiliation{University of Chinese Academy of Sciences, Beijing 100049, China}
\author{Wei Chao}
\affiliation{Center of Advanced Quantum Studies, School of Physics and Astronomy, Beijing Normal University, Beijing, 100875, China}
\author{Yu Gao}
\affiliation{State Key Laboratory of Particle Astrophysics, Institute of High Energy Physics, Chinese Academy of Sciences, Beijing 100049, China}

\begin{abstract}
Majorana fermion loops contribute a shift-symmetry violating effective potential to the coupled scalar field. When the Majorana fermion is light, a larger number of symmetry violating insertions let the loop contribution to grow increasingly infrared sensitive to the Majorana mass. With explicit calculations, we calculate the symmetry violating amplitudes, show that the scalar potential's natural scales serve as nonzero kinematic matching points which remove the superficial sensitivity to the symmetry-violating Majorana mass, continuously restoring the shift symmetry in agreement with 't Hooft naturalness. In the spontaneous broken phase, the scalar radial vacuum expectation value readily ensures the infrared safety of symmetry violating Majorana fermion loops, and all $n$-point Majoron functions develop a leading linear dependence on the symmetry violating parameter in the soft breaking limit, which matches the expansion of the Coleman-Weinberg determinant.
\end{abstract}

\maketitle

\section{Introduction}
\label{sec:intro}

The spontaneous breaking of a global $U(1)$ symmetry gives rise to a Nambu-Goldstone boson. In lepton-number and Majoron frameworks, this degree of freedom is identified with the Majoron\cite{Chikashige:1980ui,chikashige1981there,gelmini1981neutrino,Schechter:1981cv}. Majoron models have been investigated in a variety of phenomenological contexts, including dark matter scenarios~\cite{berezinsky1993majoron,GarciaCely:2017,Akita:2023,deGiorgi:2024abc,Batell:2026avi}. Recent studies have explored the interplay between Majoron dark matter and leptogenesis~\cite{King:2024idj}, as well as the prospects for detecting Majoron dark matter with axion experiments~\cite{Liang:2024vnd}. Radiative Majoron interactions are relevant for dark matter freeze-in and related cosmology~\cite{Frigerio:2011,Heeck:2017majoron,MannaSil:2023}. Mass generation for the Majoron through explicit breaking of global lepton number was studied early in connection with Planck-scale effects and cosmology~\cite{Akhmedov:1992et,Rothstein:1992rh}, while loop-induced Majoron couplings to ordinary fermions~\cite{Chang:1988dq,Pilaftsis:1993af}. Small explicit symmetry-breaking interactions, for instance through couplings to Majorana fermions, softly break the joint global \(U(1)\) shift symmetry of the coupled theory~\cite{Dashen:1971et,Weinberg:1978kz}. When the shift symmetry is slightly broken, the Majoron becomes a pseudo-Nambu--Goldstone boson, serving as a natural dark matter candidate via misalignment mechanism in the early Universe~\cite{Preskill:1983,Marsh:2016}. Radiative corrections from Majorana fermion loops then further alter the scalar potential~\cite{coleman1973radiative,Jackiw:1974cv} and are relevant across several fronts, including spontaneous
baryogenesis~\cite{Chao:2023ojl,Chun:2023eqc}, stellar cooling~\cite{Farzan:2002yp,Pilaftsis:1993af}, and neutrinoless double-beta decay~\cite{Bamert:1994qf,KamLANDZen:2020wjq}. For recent analyses of light pseudo-scalars and Majoron-like particles, see  Refs.~\cite{Arias:2012az,Dolan:2015xqa,Ballesteros:2016eoj,Rojas:2017sih,HerreroBrocal:2023xdu}.

The linear scaling of a Majoron mass is a general feature of broken global symmetries~\cite{Akhmedov:1992et,Rothstein:1992rh,Dashen:1971et}. 
However, at loop level, vacuum and zero-temperature Majorana fermion loop amplitudes rely on the Majorana mass $M$, and becomes infrared sensitive when the Majorana fermion is taken to be light. In particular, $M$ dependence worsens for the box and higher one-loop amplitudes\cite{Chao:2023ojl}. Infrared-sensitive loop amplitudes need to be matched to physical kinematical points that help alleviate unnatural dependence on the symmetry breaking parameter $M$. In a coupled theory, the theory's potential can provide such scales that regulate the amplitudes' $M$ dependence, and help to recover the symmetry in the soft-breaking limit. We consider the typical Majorana-scalar coupled theory, which incorporates the Majoron case, and perform systematic calculation of the effective scalar $n$-point amplitudes with Majorana fermion loops. These amplitudes are categorized by symmetry-breaking insertion order that provides schematic power counting and $M$-scaling relations and net phase dependencies. While the scalar $n$-point amplitudes may still be infrared sensitive, the symmetry can be shown to restore naturally.

This paper is organized as follows: Section~\ref{sec:model} discusses the zero scalar VEV scenario with a nonzero scalar mass, categorizing symmetry-breaking Majorana loops depending on their fermion number preserving (FNP) or violating (FNV) propagators, and we compute the soft-breaking behavior of the relevant amplitudes at nonzero kinematic matching points. Section~\ref{sec:theta} discusses the spontaneous symmetry broken phase where the Majoron is the Nambu--Goldstone mode, and the fermion inherits a Majorana mass from the scalar VEV. In both regimes, the explicit Majorana mass is the only source of symmetry breaking. The shift-symmetry-violating amplitudes are shown to vanish smoothly as \(M\to0\), confirming 't~Hooft naturalness~\cite{tHooft:1980,Giudice:2008bi,Penco:2020}.
The main results include the FNV/FNP power counting of symmetry violating amplitudes, the classification in terms of net phases, and the demonstration of symmetry restoration in the soft-breaking limit. 
The hierarchy between the explicit symmetry-breaking Majorana mass parameter and  other scales of the theory controls the radiative corrections. In the zero-VEV regime, the scalar mass \(m_\phi\) provides the scale required for a uniform matching expansion, ensuring that all symmetry-violating amplitudes vanish as \(M/m_\phi\to0\). In the broken phase, the soft-breaking regime \(M\ll m_0\) is regulated by \(m_0=yv/\sqrt{2}\), the fermion mass generated by the scalar vacuum expectation value.

\begin{figure}[t]
\centering
\usetikzlibrary{decorations.markings}

\tikzset{
  majorana-prop/.style={very thick},
  scalar-prop/.style={thick, dashed}
}

\begin{subfigure}[b]{0.45\textwidth}
\centering
\begin{tikzpicture}[scale=0.75]
  \useasboundingbox (-3.0, -2.5) rectangle (3.0, 2.5);
  \draw[majorana-prop] (100:1.8) arc (100:-200:1.8);
  \draw[scalar-prop] (100:1.8) -- (100:2.4) node[anchor=280, font=\small] {$\phi$};
  \draw[scalar-prop] (70:1.8) -- (70:2.4) node[anchor=250, font=\small] {$\phi$};
  \draw[scalar-prop] (40:1.8) -- (40:2.4) node[anchor=220, font=\small] {$\phi$};
  \draw[scalar-prop] (10:1.8) -- (10:2.4) node[anchor=190, font=\small] {$\phi$};
  \draw[scalar-prop] (-20:1.8) -- (-20:2.4) node[anchor=160, font=\small] {$\phi$};
  \draw[scalar-prop] (-50:1.8) -- (-50:2.4) node[anchor=130, font=\small] {$\phi$};
  \draw[scalar-prop] (-80:1.8) -- (-80:2.4) node[anchor=100, font=\small] {$\phi$};
  \draw[scalar-prop] (-110:1.8) -- (-110:2.4) node[anchor=70, font=\small] {$\phi$};
  \draw[scalar-prop] (-140:1.8) -- (-140:2.4) node[anchor=40, font=\small] {$\phi$};
  \draw[scalar-prop] (-170:1.8) -- (-170:2.4) node[anchor=10, font=\small] {$\phi$};
  \draw[scalar-prop] (-200:1.8) -- (-200:2.4) node[anchor=-20, font=\small] {$\phi$};
  \foreach \angle in {85,55,25,-5,-35,-65,-95,-125,-155,-185} {
    \draw[very thick, fill=white] (\angle:1.8) ++(-0.06,-0.06) -- ++(0.12,0.12);
    \draw[very thick, fill=white] (\angle:1.8) ++(-0.06,0.06) -- ++(0.12,-0.12);
  }
  \node[font=\Huge] at (-200:1.8) {$\cdot$};
  \node[font=\Huge] at (-220:1.8) {$\cdot$};
  \node[font=\Huge] at (-210:1.8) {$\cdot$};
  \node[font=\Huge] at (-240:1.8) {$\cdot$};
  \node[font=\Huge] at (-230:1.8) {$\cdot$};
  \node[font=\Huge] at (110:1.8) {$\cdot$};
  \node[font=\Huge] at (100:1.8) {$\cdot$};
  \node[font=\Large] at (0, 0) {$n$};
\end{tikzpicture}
\caption{Fermion number violating (with mass insertion)}
\label{fig:n-point-mass}
\end{subfigure}
\hfill
\begin{subfigure}[b]{0.45\textwidth}
\centering
\begin{tikzpicture}[scale=0.75]
  \useasboundingbox (-3.0, -2.5) rectangle (3.0, 2.5);
  \draw[majorana-prop] (100:1.8) arc (100:-200:1.8);
  \draw[scalar-prop] (100:1.8) -- (100:2.4) node[anchor=280, font=\small] {$\phi$};
  \draw[scalar-prop] (70:1.8) -- (70:2.4) node[anchor=250, font=\small] {$\phi^\ast$};
  \draw[scalar-prop] (40:1.8) -- (40:2.4) node[anchor=220, font=\small] {$\phi$};
  \draw[scalar-prop] (10:1.8) -- (10:2.4) node[anchor=190, font=\small] {$\phi^\ast$};
  \draw[scalar-prop] (-20:1.8) -- (-20:2.4) node[anchor=160, font=\small] {$\phi$};
  \draw[scalar-prop] (-50:1.8) -- (-50:2.4) node[anchor=130, font=\small] {$\phi^\ast$};
  \draw[scalar-prop] (-80:1.8) -- (-80:2.4) node[anchor=100, font=\small] {$\phi$};
  \draw[scalar-prop] (-110:1.8) -- (-110:2.4) node[anchor=70, font=\small] {$\phi^\ast$};
  \draw[scalar-prop] (-140:1.8) -- (-140:2.4) node[anchor=40, font=\small] {$\phi$};
  \draw[scalar-prop] (-170:1.8) -- (-170:2.4) node[anchor=10, font=\small] {$\phi^\ast$};
  \draw[scalar-prop] (-200:1.8) -- (-200:2.4) node[anchor=-20, font=\small] {$\phi$};
  \node[font=\Huge] at (-200:1.8) {$\cdot$};
  \node[font=\Huge] at (-220:1.8) {$\cdot$};
  \node[font=\Huge] at (-210:1.8) {$\cdot$};
  \node[font=\Huge] at (-240:1.8) {$\cdot$};
  \node[font=\Huge] at (-230:1.8) {$\cdot$};
  \node[font=\Huge] at (110:1.8) {$\cdot$};
  \node[font=\Huge] at (100:1.8) {$\cdot$};
  \node[font=\Large] at (0, 0) {$n$};
\end{tikzpicture}
\caption{FNP propagators (kinetic).}
\label{fig:n-point-no-mass}
\end{subfigure}
\caption{One-loop $n$-point diagrams with Majorana fermions. (a) FNV: mass insertions (crosses) generate operators with unequal numbers of $\phi$ and $\phi^*$.  (b) FNP: only kinetic propagators, give $U(1)$‑invariant operators.}
\label{fig:n-point-loop-comparison}
\end{figure}
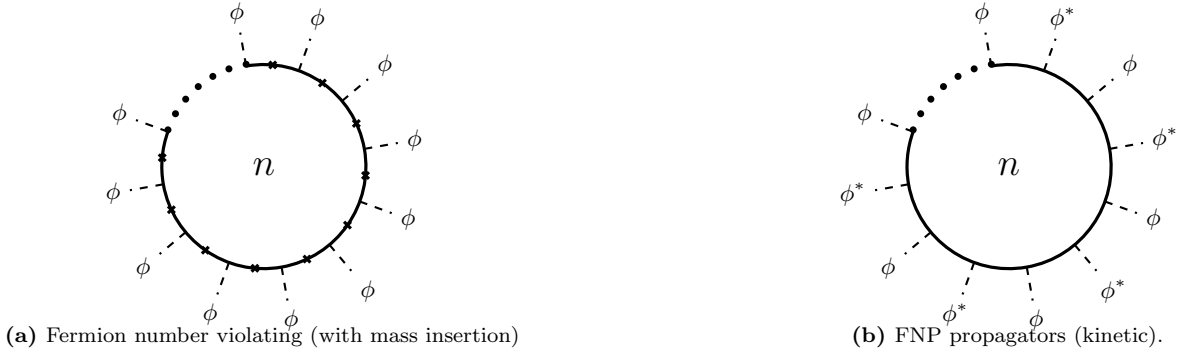

\section{Symmetry violating scalar amplitudes}
\label{sec:model}

We will first consider the scenario of coupling a perturbative scalar field to the Majorana fermion. The relevant Lagrangian for a complex scalar $\phi$ coupled to a two component Weyl Majorana fermion $\chi$ is
\begin{equation}
\mathcal{L}
=
|\partial_\mu\phi|^2
+\frac12 \chi^\dagger i\bar{\sigma}^\mu\partial_\mu\chi
+\frac12 M\chi\chi
+\frac12 y\phi\chi\chi
+\text{h.c.}
- V(|\phi|),
\label{eq:main_lagrangian}
\end{equation}
where $M$ is the explicit symmetry-breaking Majorana mass and $y$ is the Yukawa coupling. For $V(|\phi|)$ we first assume it to have a minimum at $\VEV{\phi}=0$, around which $\phi$ is perturbative. Although there is no Majoron in this zero VEV scenario, it helps clarify some infrared issues with Majorana fermion loop induced symmetry violation, in particular with the power counting of the symmetry breaking $M$ parameter in its soft limit. We will proceed with the Majoron case later in the next section. In the limit $M\to 0$, the theory respects a global symmetry,
\begin{equation}
\chi \rightarrow \chi\,e^{i\theta},
\qquad
\phi \rightarrow \phi\,e^{-i2\theta},
\qquad
\theta\in[0,2\pi),
\end{equation}
which does not depend on the existence of a scalar vacuum expectation value. $M$ is the sole source of explicit breaking of this symmetry, thus we expect all $U(1)$-violating amplitudes to vanish continuously as $M\to0$. It is interesting to see if this is the case at loop level, by examining the Majorana fermion loop amplitudes as illustrated in Fig.~\ref{fig:n-point-loop-comparison}
Symmetry violating one-loop diagrams arise from those with unequal numbers of external $\phi$ and $\phi^*$ legs, as illustrated in Fig.~\ref{fig:n-point-loop-comparison}. The Majorana fermion with $M\neq 0$ has two types of propagators, and two adjacent external scalars can be identical or conjugates depending on the type of Majorana propagator that connects them:

\begin{itemize}
    \item {Fermion-number violating (FNV)
    [Fig.~\ref{fig:n-point-mass}]:} connects two identical scalar fields
    ($\phi$--$\phi$ or $\phi^*$--$\phi^*$) via a mass insertion
    $\langle\chi\chi\rangle\propto M$, often a.k.a. the mass-inserted propagator, and is shown with a $\times$ symbol. This propagator violates fermion number by two
    units and is the source of $U(1)$ breaking.
    \item {Fermion-number preserving (FNP)
    [Fig.~\ref{fig:n-point-no-mass}]:} connects a $\phi$ and a $\phi^*$ via
    the kinetic propagator $\langle\chi\chi^\dagger\rangle\propto\sigma^\mu p_\mu$.
    This conserves fermion number and, when used alone, generates
    $U(1)$-invariant operators $|\phi|^{2n}$.
\end{itemize}
Explicitly, in two-component Weyl notation,
\begin{align}
\langle \chi_\alpha(q)\,\chi^\dagger_{\dot\beta}(-q)\rangle
&=
\frac{i(\sigma^\mu q_\mu)_{\alpha\dot\beta}}
     {q^2-M^2+i\epsilon}
\qquad \text{(FNP)},
\label{eq:FNP}
\\[2mm]
\langle \chi_\alpha(q)\,\chi_\beta(-q)\rangle
&=
\frac{iM\,\epsilon_{\alpha\beta}}
     {q^2-M^2+i\epsilon}
\qquad \text{(FNV)}.
\label{eq:FNV}
\end{align}
It is convenient to label each $n$-point one loop diagram by counting the type of internal propagators, $(N_{\rm FNV}, N_{\rm FNP})$, with $n = N_{\rm FNV}+N_{\rm FNP}$ in total. Each FNV insertion supplies one power of $M$, while FNP propagators can be momentum dominated and less sensitive to $M$. Generally, an $n$-point amplitude generates a local operator $\phi^{n-m}\phi^{*m}+\text{h.c.}$ with $0\le m\le n$. $\phi,\phi^*$ power-unmatched diagrams with $n-m\neq m$ will violate the shift symmetry. Such diagrams require at least one FNV propagator inside the loop, while the presence of FNV propagators does not necessarily guarantee symmetry violation.

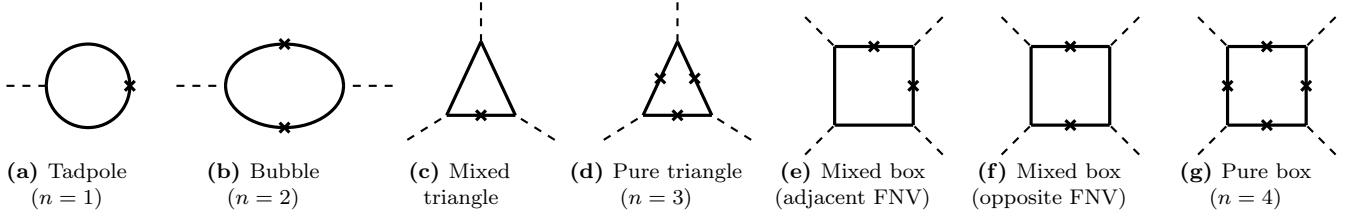
\begin{figure}[t]
\tikzset{
  majorana/.style={very thick},
  scalar/.style={thick, dashed}
}

\begin{subfigure}[b]{0.13\textwidth}
\centering
\begin{tikzpicture}[scale=0.65]
  \useasboundingbox (-2.2, -1.2) rectangle (2.2, 1.2);
  \coordinate (vL) at (-0.85, 0);
  \coordinate (vR) at (0.85, 0);
  \draw[scalar] (vL) -- (-1.7, 0);
  
  \draw[majorana] (vL) arc (180:360:0.85 and 0.85);

  \draw[majorana] (vR) arc (0:180:0.85 and 0.85);

  \draw[very thick, fill=white] (vR) ++(-0.1,-0.1) -- ++(0.2,0.2);
  \draw[very thick, fill=white] (vR) ++(-0.1,0.1) -- ++(0.2,-0.2);
\end{tikzpicture}
\caption{Tadpole ($n=1$)}
\label{fig:tadpole}
\end{subfigure}%
\hfill
\begin{subfigure}[b]{0.13\textwidth}
\centering
\begin{tikzpicture}[scale=0.65]
  \useasboundingbox (-2.2, -1.2) rectangle (2.2, 1.2);
  \coordinate (v1) at (-1.2, 0);
  \coordinate (v2) at (1.2, 0);
  \coordinate (m_top) at (0, 0.85);
  \coordinate (m_bottom) at (0, -0.85);
  \draw[scalar] (-2.2, 0) -- (v1);
  \draw[scalar] (2.2, 0) -- (v2);
  \draw[majorana] (v1) arc (180:0:1.2 and 0.85);
  \draw[majorana] (v1) arc (-180:0:1.2 and 0.85);
  \draw[very thick, fill=white] (m_top) ++(-0.1,-0.1) -- ++(0.2,0.2);
  \draw[very thick, fill=white] (m_top) ++(-0.1,0.1) -- ++(0.2,-0.2);
  \draw[very thick, fill=white] (m_bottom) ++(-0.1,-0.1) -- ++(0.2,0.2);
  \draw[very thick, fill=white] (m_bottom) ++(-0.1,0.1) -- ++(0.2,-0.2);
\end{tikzpicture}
\caption{Bubble ($n=2$)}
\end{subfigure}%
\hfill
\begin{subfigure}[b]{0.13\textwidth}
\centering
\begin{tikzpicture}[scale=0.65]
  \useasboundingbox (-2.2, -1.2) rectangle (2.2, 1.2);
  \coordinate (v1) at (-0.7, -0.6);
  \coordinate (v2) at (0.7, -0.6);
  \coordinate (v3) at (0, 0.9);
  \coordinate (m) at (0, -0.6);
  \draw[scalar] (-1.5, -1.1) -- (v1);
  \draw[scalar] (1.5, -1.1) -- (v2);
  \draw[scalar] (0, 1.7) -- (v3);
  \draw[majorana] (v1) -- (m);
  \draw[majorana] (m) -- (v2);
  \draw[majorana] (v2) -- (v3);
  \draw[majorana] (v3) -- (v1);
  \draw[very thick, fill=white] (m) ++(-0.1,-0.1) -- ++(0.2,0.2);
  \draw[very thick, fill=white] (m) ++(-0.1,0.1) -- ++(0.2,-0.2);
\end{tikzpicture}
\caption{Mixed triangle}
\end{subfigure}%
\hfill
\begin{subfigure}[b]{0.13\textwidth}
\centering
\begin{tikzpicture}[scale=0.65]
  \useasboundingbox (-2.2, -1.2) rectangle (2.2, 1.2);
  \coordinate (v1) at (-0.7, -0.6);
  \coordinate (v2) at (0.7, -0.6);
  \coordinate (v3) at (0, 0.9);
  \coordinate (m_bot) at (0, -0.6);
  \coordinate (m_right) at (0.35, 0.15);
  \coordinate (m_left) at (-0.35, 0.15);
  \draw[scalar] (-1.5, -1.1) -- (v1);
  \draw[scalar] (1.5, -1.1) -- (v2);
  \draw[scalar] (0, 1.7) -- (v3);
  \draw[majorana] (v1) -- (v2);
  \draw[majorana] (v2) -- (v3);
  \draw[majorana] (v3) -- (v1);
  \draw[very thick, fill=white] (m_bot) ++(-0.1,-0.1) -- ++(0.2,0.2);
  \draw[very thick, fill=white] (m_bot) ++(-0.1,0.1) -- ++(0.2,-0.2);
  \draw[very thick, fill=white] (m_right) ++(-0.1,-0.1) -- ++(0.2,0.2);
  \draw[very thick, fill=white] (m_right) ++(-0.1,0.1) -- ++(0.2,-0.2);
  \draw[very thick, fill=white] (m_left) ++(-0.1,-0.1) -- ++(0.2,0.2);
  \draw[very thick, fill=white] (m_left) ++(-0.1,0.1) -- ++(0.2,-0.2);
\end{tikzpicture}
\caption{Pure triangle ($n=3$)}
\end{subfigure}%
\hfill
\begin{subfigure}[b]{0.13\textwidth}
\centering
\begin{tikzpicture}[scale=0.65]
  \useasboundingbox (-2.2, -1.2) rectangle (2.2, 1.2);
  \coordinate (v1) at (-0.8, 0.8);
  \coordinate (v2) at (0.8, 0.8);
  \coordinate (v3) at (0.8, -0.8);
  \coordinate (v4) at (-0.8, -0.8);
  \coordinate (m_top) at (0, 0.8);
  \coordinate (m_right) at (0.8, 0);
  \draw[scalar] (-1.4, 1.4) -- (v1);
  \draw[scalar] (1.4, 1.4) -- (v2);
  \draw[scalar] (1.4, -1.4) -- (v3);
  \draw[scalar] (-1.4, -1.4) -- (v4);
  \draw[majorana] (v1) -- (m_top);
  \draw[majorana] (m_top) -- (v2);
  \draw[majorana] (v2) -- (m_right);
  \draw[majorana] (m_right) -- (v3);
  \draw[majorana] (v3) -- (v4);
  \draw[majorana] (v4) -- (v1);
  \draw[very thick, fill=white] (m_top) ++(-0.1,-0.1) -- ++(0.2,0.2);
  \draw[very thick, fill=white] (m_top) ++(-0.1,0.1) -- ++(0.2,-0.2);
  \draw[very thick, fill=white] (m_right) ++(-0.1,-0.1) -- ++(0.2,0.2);
  \draw[very thick, fill=white] (m_right) ++(-0.1,0.1) -- ++(0.2,-0.2);
\end{tikzpicture}
\caption{Mixed box (adjacent FNV)}
\end{subfigure}%
\hfill
\begin{subfigure}[b]{0.13\textwidth}
\centering
\begin{tikzpicture}[scale=0.65]
  \useasboundingbox (-2.2, -1.2) rectangle (2.2, 1.2);
  \coordinate (v1) at (-0.8, 0.8);
  \coordinate (v2) at (0.8, 0.8);
  \coordinate (v3) at (0.8, -0.8);
  \coordinate (v4) at (-0.8, -0.8);
  \coordinate (m_top) at (0, 0.8);
  \coordinate (m_bottom) at (0, -0.8);
  \draw[scalar] (-1.4, 1.4) -- (v1);
  \draw[scalar] (1.4, 1.4) -- (v2);
  \draw[scalar] (1.4, -1.4) -- (v3);
  \draw[scalar] (-1.4, -1.4) -- (v4);
  \draw[majorana] (v1) -- (m_top);
  \draw[majorana] (m_top) -- (v2);
  \draw[majorana] (v2) -- (v3);
  \draw[majorana] (v3) -- (m_bottom);
  \draw[majorana] (m_bottom) -- (v4);
  \draw[majorana] (v4) -- (v1);
  \draw[very thick, fill=white] (m_top) ++(-0.1,-0.1) -- ++(0.2,0.2);
  \draw[very thick, fill=white] (m_top) ++(-0.1,0.1) -- ++(0.2,-0.2);
  \draw[very thick, fill=white] (m_bottom) ++(-0.1,-0.1) -- ++(0.2,0.2);
  \draw[very thick, fill=white] (m_bottom) ++(-0.1,0.1) -- ++(0.2,-0.2);
\end{tikzpicture}
\caption{Mixed box (opposite FNV)}
\end{subfigure}%
\hfill
\begin{subfigure}[b]{0.13\textwidth}
\centering
\begin{tikzpicture}[scale=0.65]
  \useasboundingbox (-2.2, -1.2) rectangle (2.2, 1.2);
  \coordinate (v1) at (-0.8, 0.8);
  \coordinate (v2) at (0.8, 0.8);
  \coordinate (v3) at (0.8, -0.8);
  \coordinate (v4) at (-0.8, -0.8);
  \coordinate (m_top) at (0, 0.8);
  \coordinate (m_right) at (0.8, 0);
  \coordinate (m_bottom) at (0, -0.8);
  \coordinate (m_left) at (-0.8, 0);
  \draw[scalar] (-1.4, 1.4) -- (v1);
  \draw[scalar] (1.4, 1.4) -- (v2);
  \draw[scalar] (1.4, -1.4) -- (v3);
  \draw[scalar] (-1.4, -1.4) -- (v4);
  \draw[majorana] (v1) -- (m_top);
  \draw[majorana] (m_top) -- (v2);
  \draw[majorana] (v2) -- (m_right);
  \draw[majorana] (m_right) -- (v3);
  \draw[majorana] (v3) -- (m_bottom);
  \draw[majorana] (m_bottom) -- (v4);
  \draw[majorana] (v4) -- (m_left);
  \draw[majorana] (m_left) -- (v1);
  \draw[very thick, fill=white] (m_top) ++(-0.1,-0.1) -- ++(0.2,0.2);
  \draw[very thick, fill=white] (m_top) ++(-0.1,0.1) -- ++(0.2,-0.2);
  \draw[very thick, fill=white] (m_right) ++(-0.1,-0.1) -- ++(0.2,0.2);
  \draw[very thick, fill=white] (m_right) ++(-0.1,0.1) -- ++(0.2,-0.2);
  \draw[very thick, fill=white] (m_bottom) ++(-0.1,-0.1) -- ++(0.2,0.2);
  \draw[very thick, fill=white] (m_bottom) ++(-0.1,0.1) -- ++(0.2,-0.2);
  \draw[very thick, fill=white] (m_left) ++(-0.1,-0.1) -- ++(0.2,0.2);
  \draw[very thick, fill=white] (m_left) ++(-0.1,0.1) -- ++(0.2,-0.2);
\end{tikzpicture}
\caption{Pure box ($n=4$)}
\end{subfigure}
\caption{Low-order one-loop diagrams that generate symmetry-violating scalar operators (excluding sub-figure f). Thick lines: Majorana fermions; dashed lines: $\phi,\phi^*$; crosses: FNV mass insertions (each $\propto M$). (a) Tadpole: $\phi+\text{h.c.}$, removed by the choice of vacuum renormalisation condition. (b) Bubble: $\phi^2+\phi^{*2}$. (c) Mixed triangle: $\phi^2\phi^*+\text{h.c.}$. (d) Pure triangle: $\phi^3+\phi^{*3}$. (e) Mixed box with adjacent FNV insertions: $\phi^3\phi^*+\text{h.c.}$ (symmetry-violating). (f) Mixed box with opposite FNV insertions: $|\phi|^4$ ($U(1)$-invariant). (g) Pure box: $\phi^4+\phi^{*4}$.
}
\label{fig:final}
\end{figure}

A number of contributing diagrams are shown in Fig.~\ref{fig:final}, with $n$ ranging from 1 up to 4, corresponding to tadpole, bubble, triangle and box loops.
We denote diagrams with only FNV propagators as 'pure', and those containing both FNV and FNP propagators as 'mixed'.
The one-loop amplitude for a general $(N_{\rm FNV},N_{\rm FNP})$ diagram structure is schematically written as
\begin{equation}
i\mathcal{M}^{(N_{\rm FNV},N_{\rm FNP})} \sim y^n M^{N_{\rm FNV}}
\int\frac{d^4q}{(2\pi)^4}
\frac{\mathcal{N}_{N_{\rm FNP}}(q,\{p_i\})}
{\prod_{j=1}^{n}\big[(q+k_j)^2-M^2\big]},
\label{eq:zero_amp}
\end{equation}
where $M^{N_{\rm FNV}}$ denotes the number of mass-inserted fermion propagators, and $\mathcal{N}_{N_{\rm FNP}}$ denotes the  Dirac trace from the $N_{\rm FNP}$  propagators. $k_j$ ($j=0,\dots,n-1$) represent the $j$th internal propagator's accumulated input from external scalar momenta $p_i$.
For pure-FNV diagrams, the numerator is momentum-independent and the loop integral has mass dimension $4-2n$. 

For external momenta, caution is needed: it is not difficult to see Eq~\ref{eq:zero_amp} is infrared-sensitive as a lower powers of momentum are common with FNV propagators. This leads to naturalness issues if we naively assign the zero external momentum condition $p_i^2 =0$, in which $M$ becomes the only dimensionful parameter, and pure-FNV box amplitude ($N_{\rm FNV}=4$) evaluates to an $M$-independent constant~\cite{Chao:2023ojl}, and all $N_{\rm FNV}>4$ diagrams are inversely divergent as $M\rightarrow 0$. This contradiction with symmetry restoration toward zero $M$ means the external $p_i^2=0$ is not an optimal choice for these amplitudes, and we need to consider some other kinematic points. Note that the scalar potential shape at its $\VEV{\phi}=0$ minimum conveniently provides the scale mass $m_\phi$ as a reference scale. For $M\rightarrow 0$, we will consider a hierarchical kinematic region $|\vec{k}|\ll M \ll m_\phi$, and evaluate the amplitudes for $p_i=(p_i^0,\vec{0})$. For convenience, we take $p^0_i=m_\phi$ that represents the on-shell amplitudes for an even number of external scalars. For diagrams with an odd number of the external scalars, we take some of the $p_i$ to be virtual to satisfy momentum conservation. Such kinematic assignments do not remove infrared sensitivity; the amplitudes in Eq.~\ref{eq:zero_amp} rely on the external momenta and some nonzero assignment is needed to avoid divergence. However, we will show that with the help of this hierarchical assignment, the global shift symmetry is restored in the $M/m_\phi \rightarrow 0$ limit.

For $n>2$ pure-FNV diagrams the integrals are UV-finite; with $p_i^2\sim m_\phi^2$, the amplitude follows the scaling relation
\begin{equation}
i\mathcal{M}_n \sim y^n M^n \, m_\phi^{4-2n}.
\label{eq:pure_scaling_fixed}
\end{equation}
The $n=2$ bubble amplitude is logarithmically divergent with a prefactor proportional to $M^2$, a pattern that extends to the pure triangle and pure box, which scale as $M^3$ and $M^4$, respectively. To demonstrate the general power-counting rules and to identify the $M$ power counting of the symmetry violation, we now calculate the pure-FNV and mixed diagrams up to $n=4$ and find that the pure-FNV bubble, triangle, and box generate $\phi^2$, $\phi^3$, and $\phi^4$, respectively; the mixed triangle yields $\phi^2\phi^*+\mathrm{h.c.}$, while the mixed box splits into an adjacent configuration producing the symmetry-violating $\phi^3\phi^*+\mathrm{h.c.}$ and an opposite configuration producing the $U(1)$-invariant $|\phi|^4$ (see Fig.~\ref{fig:final}).

\medskip

The tadpole ($N_{\rm FNV}=1$) is removed by requiring $\langle\phi\rangle=0$ as the physical vacuum. This renormalization condition 
defines the perturbative expansion point around the symmetric vacuum~\cite{Masina:2025}. All physical symmetry‑breaking observables remain proportional to powers of the explicit mass $M$. In the FNV/FNP classification this first-order explicit-breaking contribution corresponds to the mixed triangle $(1,2)$, whose spinor trace supplies a $q^2$ numerator that raises the superficial degree of divergence to $D=0$; it therefore does not follow the pure-FNV scaling of Eq.~\eqref{eq:pure_scaling_fixed}. Higher-point diagrams are suppressed as higher-dimensional operators, with finite logarithmic dependence determined by the full on-shell amplitude integration. To keep the discussion concise, we only list the loop amplitude results here; loop  calculation follows standard routes and the details are available in Appendix~\ref{app:loops}.

\medskip
The pure bubble diagram has two external $\phi$ legs and two FNV propagators, generating a symmetry-violating diagram. Applying the Majorana Feynman rules~\cite{Denner:1992vza,Dreiner:2008tw}, the bubble amplitude is
\begin{equation}
i\mathcal{M}_2 = (iy)^2 \int\!\frac{d^4q}{(2\pi)^4}
\frac{(iM)^2\,\epsilon^{\alpha\beta}\epsilon_{\alpha\beta}}
{(q^2-M^2)\,[(q+p_1)^2-M^2]}
= -2y^2M^2 \, I_2(p_1^2),
\end{equation}
with
\begin{equation}
I_2(p_1^2) = \int\!\frac{d^4q}{(2\pi)^4}
\frac{1}{(q^2-M^2)\,[(q+p_1)^2-M^2]} .
\end{equation}
Evaluating the two-point function on shell at $p_1^2=m_\phi^2$ and using dimensional regularisation with $\overline{\rm MS}$ subtraction~\cite{tHooft:1978jhc,Passarino:1978jh}, one obtains
\begin{equation}
I_2^{\overline{\text{MS}}}(m_\phi^2) = \frac{i}{16\pi^2}
\Bigl( -\ln\frac{m_\phi^2}{\mu^2} + 2 \Bigr)
+ \mathcal{O}\!\left(\frac{M^2}{m_\phi^2}\right),
\end{equation}
where the constant $2$ is scheme-dependent. The renormalised amplitude and the resulting effective Lagrangian coefficient are therefore
\begin{align}
i\mathcal{M}_2^{\text{ren}}
&= \frac{i y^2 M^2}{8\pi^2} \Bigl( \ln\frac{m_\phi^2}{\mu^2} - 2 \Bigr), \\
\lambda_2
&= \frac{y^2 M^2}{16\pi^2}\left(\ln\frac{m_\phi^2}{\mu^2} - 2\right),
\label{eq:bubble_zero}
\end{align}
with $\mathcal{L}_{\text{eff}} \supset -\lambda_2 (\phi^2+\phi^{*2})$, which scales as $M^2$ with a scheme-dependent constant that does not affect the qualitative conclusion.

\medskip

The pure triangle $(N_{\rm FNV},N_{\rm FNP})=(3,0)$ diagram comprises of three external $\phi$ legs and three FNV propagators. To conserve the total momentum , so a real fully on-shell three-point kinematic configuration does not exist for identical massive scalars, this contribution is best interpreted as an off-shell three-point, evaluated in the kinematic configuration $p_1=-p_2$, $p_3=0$ with $p^2=m_\phi^2$. The amplitude is
\begin{equation}
i\mathcal M_3
=
-\frac{i\,y^3M^3}{16\pi^2}
\int_0^1 dx\,
\frac{x}{M^2-x(1-x)m_\phi^2-i\epsilon},
\label{eq:pure_tri_amplitude}
\end{equation}
where the scalar integral is UV finite ($D=-2$). In the regime $M^2\ll m_\phi^2$, the amplitude scales as
\begin{equation}
\mathcal M_3
\sim
\frac{y^3M^3}{16\pi^2m_\phi^2}
\ln\frac{m_\phi^2}{M^2}.
\end{equation}
Thus the pure-FNV triangle is protected against infrared enhancement and vanishes smoothly as $M\to0$.

\begin{figure}[ht]
    \centering
    \includegraphics[width=0.44\linewidth]{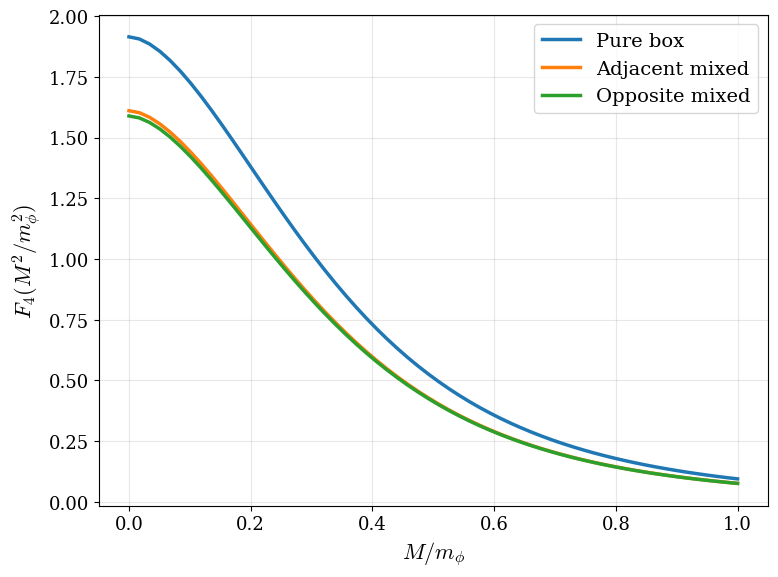}
    \includegraphics[width=0.44\linewidth]{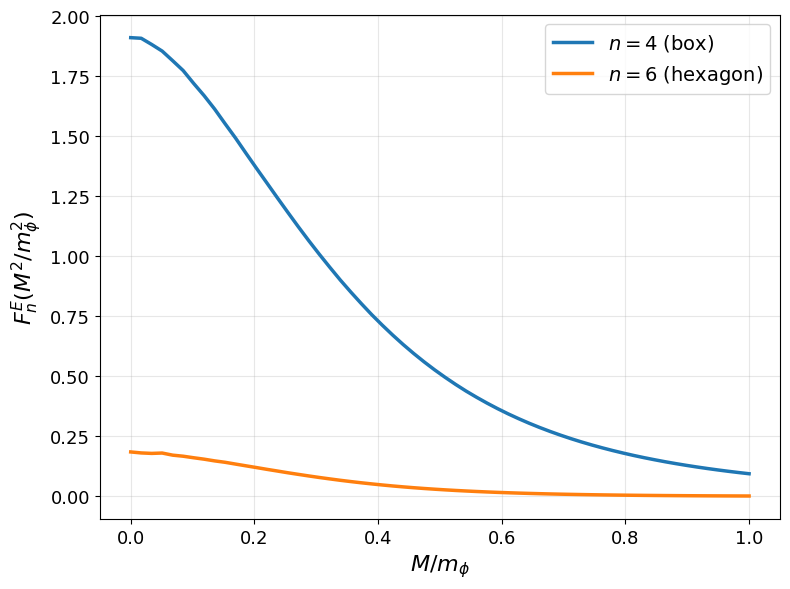}
    \caption{(left) $F_4$ functions for the three box diagrams shown in Fig.~\ref{fig:final}e-\ref{fig:final}f. With the symmetric on-shell assignment $p_i^2=m_\phi^2$, $p_i\cdot p_j=-m_\phi^2/3$, the two mixed-box functions converge. All functions have finite zero-point values: $F_{4,\mathrm{pure}}^{E}(0)=1.9$, $F_{4,\mathrm{adj}}^{E}(0), F_{4,\mathrm{opp}}^{E}(0)=1.6$.
    (right)
    comparing pure‑FNV loop functions for the box ($n=4$), and the hexagon ($n=6$) diagrams.
    All functions remain finite as $M/m_\phi\to0$.  The numerical zero limits are $F_4^{E}(0)=1.9$, $F_6^{E}(0)=0.18$.
    }
    \label{fig:integrals}
\end{figure}

For the mixed triangle $(N_{\rm FNV},N_{\rm FNP})=(1,2)$ diagram,
evaluated in the same kinematic configuration as the pure triangle
($p_1=-p_2$, $p_3=0$ with $p^2=m_\phi^2$),  the amplitude is
\begin{equation}
i\mathcal M_{3}^{\rm mix}
=
y^3M\, I_{\rm mix}(m_\phi^2),
\end{equation}
with
\begin{equation}
I_{\rm mix}(m_\phi^2)
=
\frac{i}{8\pi^2}
\left[
1-\ln\frac{m_\phi^2}{\mu^2}
+i\pi
+
\mathcal O\!\left(
\frac{M^2}{m_\phi^2}
\ln\frac{M^2}{m_\phi^2}
\right)
\right].
\end{equation}
The amplitude scales linearly in explicit breaking mass M, i.e. $\mathcal M_3^{\rm mix} = \mathcal O(y^3M)$, which vanishes as $M\to0$. Unlike the pure triangle, the
mixed triangle is logarithmically ultraviolet divergent and requires a counter-term. The explicit calculation is presented in Appendix~\ref{app:loops}.

The pure box ($N_{\rm FNV}=4$) diagram has four external $\phi$ legs and four FNV propagators. Taking external $\vec{p}_\phi=0$, the ultraviolet finite scalar four-point function can be written as
\begin{equation}
I_4(m_\phi^2) \equiv \int\!\frac{d^d q}{(2\pi)^d}
\frac{1}{\prod_{j=1}^{4}\big[(q+k_j)^2-M^2\big]}
= \frac{i}{16\pi^2}\,\frac{1}{m_\phi^4}\,
F_4\!\left(\frac{M^2}{m_\phi^2}\right),
\label{eq:F4_def}
\end{equation}
where $F_4(r)$ is a finite kinematic function. The corresponding amplitude is
\begin{equation}
i\mathcal{M}_4 = \,2 y^4 M^4 I_4(m_\phi^2)
\simeq \frac{2\,i\,y^4 M^4}{16\pi^2 m_\phi^4}\,
F_4\!\left(\frac{M^2}{m_\phi^2}\right)
=
\frac{i\,y^4 M^4}{8\pi^2 m_\phi^4}\,
F_4\!\left(\frac{M^2}{m_\phi^2}\right).
\end{equation}
In contrast to the bubble and mixed triangle, the pure box is suppressed by $M^4/m_\phi^4$ and is independent of the UV cutoff. Appendix~\ref{app:box} contains detailed information about the kinematic function. A low-$r$ expansion describes the behavior in the soft breaking limit for $r\equiv M^2/m_\phi^2 \ll 1$. The finite limit of the Feynman-parameter integral at $r=0$ implies that there are no infrared-enhanced inverse powers of $r$. The leading non-analytic correction typically arises from endpoint regions of parameter space and takes the generic form $r\ln r$:
\begin{equation}
F_4(r)=F_4(0)+A~r\ln r + \cdots, \qquad r\rightarrow 0,
\end{equation}
with nonzero
\begin{equation}
F_4(0)=
\int d\Pi_4
\frac{1}{(f_4(x_i))^2} \approx 1.9~\text{ and }~ A\approx 3.1,
\end{equation}
where the finite Feynman-parameter integral is directly evaluated numerically to yield \(F_4(0)\) and a numerical fit is used to determine \(A\). For higher $n$-point pure FNV kinematic functions, we show the hexagon integral in the $0<r<1$ range in Fig.~\ref{fig:integrals}, along with the pure FNV $F_4$ function with the same kinematic assignments.

\medskip

The mixed boxes $(N_{\rm FNV},N_{\rm FNP})=(2,2)$ have two different configurations, which depends on the FNV propagator positioning along the loop. To observe the configuration first comes with opposite insertions (Fig.~\ref{fig:final}(f)), that yields the $U(1)$-invariant operator $|\phi|^4$, that preserve the symmetry, while the symmetry-violating operator $\phi^3\phi^*+\mathrm{h.c.}$ is produced by the configuration with adjacent insertions (Fig.~\ref{fig:final}(e)). See Appendix~\ref{app:box} for computation details.

\medskip

For a mini-summary, in this section we showed that
every symmetry-violating loop diagram contains at least one FNV insertion and thus at least one power of \(M\), but their amplitudes face an infrared sensitivity issue at low $M$, especially if $M$ appears as the only nonzero scale of the calculation. A nonzero external momentum assignment will protect the amplitudes from developing dangerous $M^{4-n}$ dependence as in the zero external momentum point. We explicitly show that the symmetry-breaking bubble, triangle, and box amplitudes scale to zero with $M$ in the $|\vec{k}|\ll M \ll m_\phi$ regime. $m_\phi>0$ at the $V(|\phi|)$ minimum is a natural choice and plays the role of a regulatory external momentum scale. This restores the symmetry and confirms 't~Hooft naturalness~\cite{Penco:2020,Giudice:2008bi}. 
In case of a significant $M$, these diagrams contribute to symmetry violating correlations such as $\{\phi^n+\phi^{\ast n}\}$ that would distort the shape of the effective scalar potential. In the next section, we consider the spontaneously broken \(U(1)\) phase, where the Majoron potential encodes the same explicit shift-symmetry breaking, and re-derive the Majorana-loop induced contribution in the soft-breaking (low-\(M\)) limit.

\section{Majoron potential from Majorana fermion loop}
\label{sec:theta}

We now consider the spontaneously broken phase. The scalar field develops a radial VEV $\VEV{|\phi|}\neq 0$, and the pseudo NG mode is the usual Majoron.
The effective theory is expanded around the broken vacuum, defining a different effective description from the zero‑VEV on‑shell amplitudes of Sec.~\ref{sec:model}.
In the phase-shift direction, the periodic structure of the resulting Majoron potential emerges after projecting the Nambu-Goldstone dependent effective action onto the zero‑momentum non‑derivative sector~\cite{coleman1973radiative,Jackiw:1974cv}. For the Majorana fermion, the $(N_{\rm FNV},N_{\rm FNP})$ classification of Sec.~\ref{sec:model} still applies to loop contributions according to their fermion‑number violating and preserving propagator structure. The main difference from the $v_\phi\neq 0$ case is that a redefinition of the fermion field (i.e. chiral rotation in chiral theories) can absorb the scalar, and the rotated fermion field acquires a Majorana mass from the scalar VEV, $m_0 = yv/\sqrt{2}$.
This effective mass does not break the shift‑symmetry, and it enters loop kinematics as an existing scale of the theory. The only explicit breaking still occurs with the non‑rotated fermion's Majorana mass parameter $M$, which we assume to be present, and discuss its role in generating the Majoron potential. In the soft explicit breaking limit, we assume $p_\theta\to0$ and $M\ll m_0$.

\begin{figure}[H]
\centering
\tikzset{
  majorana/.style={very thick},
  scalar/.style={thick, dashed}
}

\begin{subfigure}[b]{0.17\textwidth}
\centering
\begin{tikzpicture}[scale=0.65]
  \useasboundingbox (-2.2, -1.2) rectangle (2.2, 1.2);
  \coordinate (vB) at (0, -0.85);
  \coordinate (vT) at (0, 0.85);

  \draw[scalar] (-1.2, -1.4) -- (vB);
  \draw[scalar] (1.2, -1.4) -- (vB);

  \draw[majorana] (vB) arc (270:450:0.85 and 0.85);
  \draw[majorana] (vT) arc (90:270:0.85 and 0.85);

  \draw[very thick] (vT) ++(-0.1,-0.1) -- ++(0.2,0.2);
  \draw[very thick] (vT) ++(-0.1,0.1) -- ++(0.2,-0.2);
\end{tikzpicture}
\caption{Tadpole (2-point)}
\label{fig:single_insertion}
\end{subfigure}%
\hfill
\begin{subfigure}[b]{0.17\textwidth}
\centering
\begin{tikzpicture}[scale=0.65]
  \useasboundingbox (-2.2, -1.2) rectangle (2.2, 1.2);
  \coordinate (v1) at (-1.2, 0);
  \coordinate (v2) at (1.2, 0);
  \coordinate (m_top) at (0, 0.85);
  \coordinate (m_bottom) at (0, -0.85);

  \draw[scalar] (-2.2, 0) -- (v1);
  \draw[scalar] (2.2, 0) -- (v2);

  \draw[majorana] (v1) arc (180:0:1.2 and 0.85);
  \draw[majorana] (v1) arc (-180:0:1.2 and 0.85);

  \draw[very thick] (m_top) ++(-0.1,-0.1) -- ++(0.2,0.2);
  \draw[very thick] (m_top) ++(-0.1,0.1) -- ++(0.2,-0.2);
  \draw[very thick] (m_bottom) ++(-0.1,-0.1) -- ++(0.2,0.2);
  \draw[very thick] (m_bottom) ++(-0.1,0.1) -- ++(0.2,-0.2);
\end{tikzpicture}
\caption{Bubble}
\label{fig:bubble}
\end{subfigure}%
\hfill
\begin{subfigure}[b]{0.17\textwidth}
\centering
\begin{tikzpicture}[scale=0.65]
  \useasboundingbox (-2.2, -1.2) rectangle (2.2, 1.2);

  \coordinate (vL) at (-1.2, 0);
  \coordinate (vR) at (1.2, 0);
  \coordinate (m_top) at (0, 0.85);
  \coordinate (m_bot) at (0, -0.85);

  \draw[scalar] (-2.2, 0.9) -- (vL);
  \draw[scalar] (-2.4, 0.0) -- (vL);
  \draw[scalar] (-2.2, -0.9) -- (vL);

  \draw[majorana] (vL) arc (180:0:1.2 and 0.85);
  \draw[majorana] (vL) arc (-180:0:1.2 and 0.85);

  \draw[very thick] (vR) ++(-0.1,-0.1) -- ++(0.2,0.2);
  \draw[very thick] (vR) ++(-0.1,0.1) -- ++(0.2,-0.2);
\end{tikzpicture}
\caption{Tadpole (3-point)}
\label{fig:3-0}
\end{subfigure}%
\hfill
\begin{subfigure}[b]{0.17\textwidth}
\centering
\begin{tikzpicture}[scale=0.65]
  \useasboundingbox (-2.2, -1.2) rectangle (2.2, 1.2);

  \coordinate (vL) at (-1.2, 0);
  \coordinate (vR) at (1.2, 0);
  \coordinate (m_top) at (0, 0.85);
  \coordinate (m_bot) at (0, -0.85);

  \draw[scalar] (-2.2, 0.9) -- (vL);
  \draw[scalar] (-2.2, -0.9) -- (vL);
  \draw[scalar] (2.2, 0) -- (vR);

  \draw[majorana] (vL) arc (180:0:1.2 and 0.85);
  \draw[majorana] (vL) arc (-180:0:1.2 and 0.85);

  \draw[very thick] (m_top) ++(-0.1,-0.1) -- ++(0.2,0.2);
  \draw[very thick] (m_top) ++(-0.1,0.1) -- ++(0.2,-0.2);
  \draw[very thick] (m_bot) ++(-0.1,-0.1) -- ++(0.2,0.2);
  \draw[very thick] (m_bot) ++(-0.1,0.1) -- ++(0.2,-0.2);
\end{tikzpicture}
\caption{Bubble (3-point, non-uniform)}
\label{fig:2-1}
\end{subfigure}%
\hfill
\begin{subfigure}[b]{0.15\textwidth}
\centering
\begin{tikzpicture}[scale=0.65]
  \useasboundingbox (-2.2, -1.2) rectangle (2.2, 1.7);

  \coordinate (v1) at (-0.9, -0.65);
  \coordinate (v2) at (0.9, -0.65);
  \coordinate (v3) at (0, 0.95);

  \coordinate (m12) at (0, -0.65);
  \coordinate (m23) at (0.45, 0.15);
  \coordinate (m31) at (-0.45, 0.15);

  \draw[scalar] (-1.6, -1.25) -- (v1);
  \draw[scalar] (1.6, -1.25) -- (v2);
  \draw[scalar] (0, 1.65) -- (v3);

  \draw[majorana] (v1) -- (m12); \draw[majorana] (m12) -- (v2);
  \draw[majorana] (v2) -- (m23); \draw[majorana] (m23) -- (v3);
  \draw[majorana] (v3) -- (m31); \draw[majorana] (m31) -- (v1);

  \draw[very thick] (m12) ++(-0.1,-0.1) -- ++(0.2,0.2);
  \draw[very thick] (m12) ++(-0.1,0.1) -- ++(0.2,-0.2);

  \draw[very thick] (m23) ++(-0.1,-0.1) -- ++(0.2,0.2);
  \draw[very thick] (m23) ++(-0.1,0.1) -- ++(0.2,-0.2);

  \draw[very thick] (m31) ++(-0.1,-0.1) -- ++(0.2,0.2);
  \draw[very thick] (m31) ++(-0.1,0.1) -- ++(0.2,-0.2);
\end{tikzpicture}
\caption{triangle}
\label{fig:pure_triangle}
\end{subfigure}%

\hfill
\begin{subfigure}[b]{0.15\textwidth}
\centering
\begin{tikzpicture}[scale=0.65]
  \useasboundingbox (-2.2, -1.2) rectangle (2.2, 1.2);

  \coordinate (vL) at (-1.2, 0);
  \coordinate (vR) at (1.2, 0);
  \coordinate (m_top) at (0, 0.85);
  \coordinate (m_bot) at (0, -0.85);

  \draw[scalar] (-2.2, 0.9) -- (vL);
  \draw[scalar] (-2.5, 0.4) -- (vL);
  \draw[scalar] (-2.5, -0.4) -- (vL);
  \draw[scalar] (-2.2, -0.9) -- (vL);

  \draw[majorana] (vL) arc (180:0:1.2 and 0.85);
  \draw[majorana] (vL) arc (-180:0:1.2 and 0.85);

  \draw[very thick] (vR) ++(-0.1,-0.1) -- ++(0.2,0.2);
  \draw[very thick] (vR) ++(-0.1,0.1) -- ++(0.2,-0.2);
\end{tikzpicture}
\caption{Tadpole (4-point)}
\label{fig:4-0}
\end{subfigure}%
\hfill
\begin{subfigure}[b]{0.17\textwidth}
\centering
\begin{tikzpicture}[scale=0.65]
  \useasboundingbox (-2.2, -1.2) rectangle (2.2, 1.2);

  \coordinate (vL) at (-1.2, 0);
  \coordinate (vR) at (1.2, 0);
  \coordinate (m_top) at (0, 0.85);
  \coordinate (m_bot) at (0, -0.85);

  \draw[scalar] (-2.2, 0.9) -- (vL);
  \draw[scalar] (-2.4, 0.0) -- (vL);
  \draw[scalar] (-2.2, -0.9) -- (vL);
  \draw[scalar] (2.2, 0) -- (vR);

  \draw[majorana] (vL) arc (180:0:1.2 and 0.85);
  \draw[majorana] (vL) arc (-180:0:1.2 and 0.85);

  \draw[very thick] (m_top) ++(-0.1,-0.1) -- ++(0.2,0.2);
  \draw[very thick] (m_top) ++(-0.1,0.1) -- ++(0.2,-0.2);
  \draw[very thick] (m_bot) ++(-0.1,-0.1) -- ++(0.2,0.2);
  \draw[very thick] (m_bot) ++(-0.1,0.1) -- ++(0.2,-0.2);
\end{tikzpicture}
\caption{Bubble (4-point, non-uniform)}
\label{fig:3-1}
\end{subfigure}%
\hfill
\begin{subfigure}[b]{0.19\textwidth}
\centering
\begin{tikzpicture}[scale=0.65]
  \useasboundingbox (-2.2, -1.2) rectangle (2.2, 1.2);

  \coordinate (vL) at (-1.2, 0);
  \coordinate (vR) at (1.2, 0);
  \coordinate (m_top) at (0, 0.85);
  \coordinate (m_bot) at (0, -0.85);

  \draw[scalar] (-2.2, 0.9) -- (vL);
  \draw[scalar] (-2.2, -0.9) -- (vL);
  \draw[scalar] (2.2, 0.9) -- (vR);
  \draw[scalar] (2.2, -0.9) -- (vR);

  \draw[majorana] (vL) arc (180:0:1.2 and 0.85);
  \draw[majorana] (vL) arc (-180:0:1.2 and 0.85);

  \draw[very thick] (m_top) ++(-0.1,-0.1) -- ++(0.2,0.2);
  \draw[very thick] (m_top) ++(-0.1,0.1) -- ++(0.2,-0.2);
  \draw[very thick] (m_bot) ++(-0.1,-0.1) -- ++(0.2,0.2);
  \draw[very thick] (m_bot) ++(-0.1,0.1) -- ++(0.2,-0.2);
\end{tikzpicture}
\caption{Bubble (4-point, uniform)}
\label{fig:2-2}
\end{subfigure}
\hfill
\begin{subfigure}[b]{0.19\textwidth}
\centering
\begin{tikzpicture}[scale=0.65]
  \useasboundingbox (-2.2, -1.2) rectangle (2.2, 1.7);
  \coordinate (v1) at (-0.9, -0.65);
  \coordinate (v2) at (0.9, -0.65);
  \coordinate (v3) at (0, 0.95);
  \coordinate (m12) at (0, -0.65);
  \coordinate (m23) at (0.45, 0.15);
  \coordinate (m31) at (-0.45, 0.15);

  \draw[scalar] (-1.6, -1.25) -- (v1);
  \draw[scalar] (-1.6, -0.1) -- (v1);

  \draw[scalar] (1.6, -1.25) -- (v2);
 
  \draw[scalar] (0, 1.65) -- (v3);

  \draw[majorana] (v1) -- (m12); \draw[majorana] (m12) -- (v2);
  \draw[majorana] (v2) -- (m23); \draw[majorana] (m23) -- (v3);
  \draw[majorana] (v3) -- (m31); \draw[majorana] (m31) -- (v1);

  \draw[very thick] (m12) ++(-0.1,-0.1) -- ++(0.2,0.2);
  \draw[very thick] (m12) ++(-0.1,0.1) -- ++(0.2,-0.2);

  \draw[very thick] (m23) ++(-0.1,-0.1) -- ++(0.2,0.2);
  \draw[very thick] (m23) ++(-0.1,0.1) -- ++(0.2,-0.2);

  \draw[very thick] (m31) ++(-0.1,-0.1) -- ++(0.2,0.2);
  \draw[very thick] (m31) ++(-0.1,0.1) -- ++(0.2,-0.2);
\end{tikzpicture}
\caption{Triangle (4-point, non-uniform)}
\label{fig:nonuniform_triangle}
\end{subfigure}
\hfill
\begin{subfigure}[b]{0.15\textwidth}
\centering
\begin{tikzpicture}[scale=0.65]
  \useasboundingbox (-2.2, -1.2) rectangle (2.2, 1.2);

  \coordinate (v1) at (-0.8, 0.8);
  \coordinate (v2) at (0.8, 0.8);
  \coordinate (v3) at (0.8, -0.8);
  \coordinate (v4) at (-0.8, -0.8);

  \coordinate (m1) at (0, 0.8);
  \coordinate (m2) at (0.8, 0);
  \coordinate (m3) at (0, -0.8);
  \coordinate (m4) at (-0.8, 0);

  \draw[scalar] (-1.4, 1.4) -- (v1);
  \draw[scalar] (1.4, 1.4) -- (v2);
  \draw[scalar] (1.4, -1.4) -- (v3);
  \draw[scalar] (-1.4, -1.4) -- (v4);

  \draw[majorana] (v1) -- (m1); \draw[majorana] (m1) -- (v2);
  \draw[majorana] (v2) -- (m2); \draw[majorana] (m2) -- (v3);
  \draw[majorana] (v3) -- (m3); \draw[majorana] (m3) -- (v4);
  \draw[majorana] (v4) -- (m4); \draw[majorana] (m4) -- (v1);

  \draw[very thick] (m1) ++(-0.1,-0.1) -- ++(0.2,0.2);
  \draw[very thick] (m1) ++(-0.1,0.1) -- ++(0.2,-0.2);

  \draw[very thick] (m2) ++(-0.1,-0.1) -- ++(0.2,0.2);
  \draw[very thick] (m2) ++(-0.1,0.1) -- ++(0.2,-0.2);

  \draw[very thick] (m3) ++(-0.1,-0.1) -- ++(0.2,0.2);
  \draw[very thick] (m3) ++(-0.1,0.1) -- ++(0.2,-0.2);

  \draw[very thick] (m4) ++(-0.1,-0.1) -- ++(0.2,0.2);
  \draw[very thick] (m4) ++(-0.1,0.1) -- ++(0.2,-0.2);
\end{tikzpicture}
\caption{box}
\label{fig:non_deriv_box}
\end{subfigure}%
\caption{
One-loop $n=2,3,4$ point contributions to the broken-phase Majoron effective action. (a) Tadpole (2-point) diagram with
the local $a^2\chi\chi$ vertex, giving the leading contribution to Majoron mass. (b) two point function with two vertices.
(c-e) three point functions; (f-j) four point functions. Every vertex carries an $M$ factor and the diagrams for each $n$ are arranged in the rising number of vertices.
}
\label{fig:goldstone_diagrams}
\end{figure}
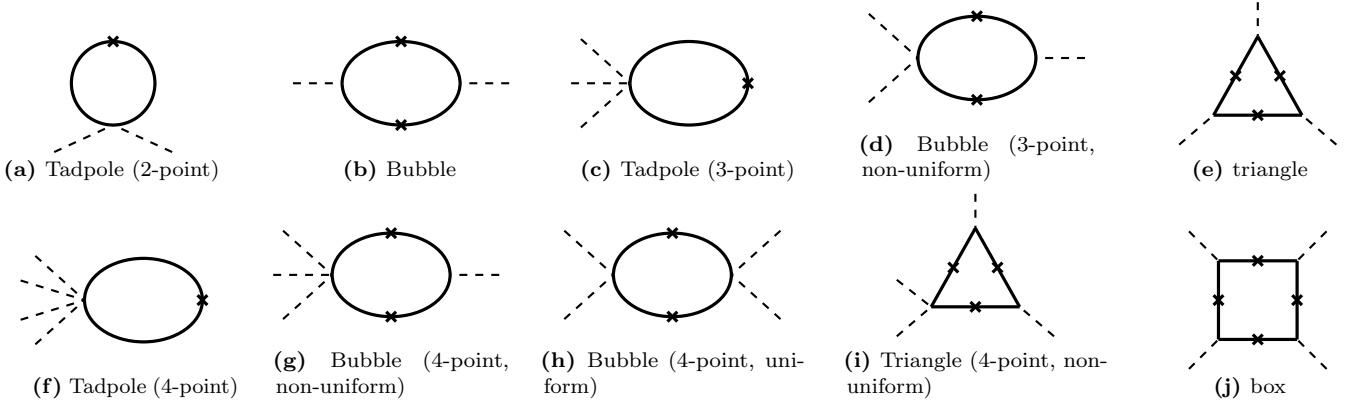

We parametrize the complex scalar in the usual way
\begin{equation}
    \phi(x)=\frac{v+\rho(x)}{\sqrt{2}}\,e^{i\theta(x)},
\label{eq:phi_param}
\end{equation}
with a heavy radial mode $\rho(x)$ and dimensionless compact field $\theta(x)$. At energies below the radial VEV, $\theta(x)$ is the only light scalar degree of freedom~\cite{Coleman:1969sm,Callan:1969sn}, and the canonically normalised Majoron field is $a(x)=v\,\theta(x)$. The coupling term now contains the phase,
\begin{equation}
\mathcal{L}_m = -\frac12\bigl(m_0 e^{i\theta} + M\bigr)\chi\chi + \mathrm{h.c.},
\qquad m_0 \equiv \frac{y v}{\sqrt{2}} \gg M,
\end{equation}
and we perform the nonlinear redefinition $\chi=e^{-i\theta/2}\tilde{\chi},$ which makes the realization of the spontaneously broken symmetry manifest. Since the Majorana bilinear transforms as $\chi\chi=e^{-i\theta}\tilde{\chi}\tilde{\chi},$ these terms become
\begin{equation}
\begin{aligned}
\mathcal{L}_{m}
&=
-\frac12
(m_0 e^{i\theta}+M)e^{-i\theta}
\tilde{\chi}\tilde{\chi}
+\mathrm{h.c.}
=
-\frac12 m_0\,\tilde{\chi}\tilde{\chi}
-\frac12 M e^{-i\theta}\tilde{\chi}\tilde{\chi}
+\mathrm{h.c.}
\end{aligned}
\label{eq:broken_mass}
\end{equation}
where the first term defines the spontaneous Majorana mass and hence the fermion propagator in the broken phase. This conveniently removes the phase from the leading spontaneous mass term while retaining the dynamical Majoron degree of freedom. The Nambu-Goldstone dependence is transferred from the mass term into derivative interactions generated by the fermion kinetic term, as required by the nonlinear realization of the broken symmetry. The non-derivative Majoron potential originates exclusively from the explicit-breaking term proportional to \(M\), which retains the physical phase dependence. For notation simplicity, we subsequently relabel \(\tilde{\chi}\rightarrow\chi\) in the rest of this section. The low-energy effective Lagrangian after redefinition is
\begin{equation}
\begin{aligned}
\mathcal{L}_{\rm broken}
&=
\frac{v^{2}}{2}(\partial_{\mu}\theta)^{2}
+\frac12\chi^{\dagger}i\bar{\sigma}^{\mu}\partial_{\mu}\chi
-\frac12 m_{0}\chi\chi
-\frac12 m_{0}\chi^{\dagger}\chi^{\dagger}
\\
&\quad
+\frac12(\partial_{\mu}\theta)
\chi^{\dagger}\bar{\sigma}^{\mu}\chi
-\frac12 M e^{-i\theta}\chi\chi
-\frac12 M e^{i\theta}\chi^{\dagger}\chi^{\dagger}.
\label{eq:Leff_theta}
\end{aligned}
\end{equation}
The $a(x)$ field remains a propagating degree of freedom in the broken phase. Its derivative coupling to the fermion 
$
\mathcal{L}_{\partial}
=
\frac{1}{2v}
(\partial_\mu a)
\chi^\dagger\bar{\sigma}^{\mu}\chi
$ emerges as a consequence of field redefinition, but makes no contribution to the Majoron potential as it is shift-symmetric. Nevertheless, it contributes to the kinetic normalization and higher-derivative structure of the full Majoron effective action.
The explicit shift-symmetry breaking is $\theta$-dependent, and we expect its vacuum value $\theta_0$ to be defined as the minimum of the non-derivative effective action $V_{\rm eff}(\theta)$:
\begin{equation}
\left.
\frac{\partial V_{\rm eff}}{\partial\theta}
\right|_{\theta=\theta_0}
=0,
\qquad
\left.
\frac{\partial^2 V_{\rm eff}}{\partial\theta^2}
\right|_{\theta=\theta_0}
>0.
\label{eq:tadpole_cond}
\end{equation}
In the low-temperature limit we can perturbatively expand the field around its minimum.
Thus, the perturbative explicit-breaking interaction expands as
\begin{equation}
\mathcal{L}_{\rm br}
=
-\frac12 M e^{-i\theta_0}
\left(
1-\frac{i a}{v}
-\frac{a^2}{2v^2}
+\mathcal O(a^3)
\right)\chi\chi
+\mathrm{h.c.}
\label{eq:vertex_tower_local}
\end{equation}
For an $a^n\chi\chi$ vertex, the identical-field contractions
cancel the $1/(2\,n!)$ in the Lagrangian coefficient, so the
Feynman rules for these vertices are
$V_n = -i M e^{-i\theta_0}(-i/v)^n$.
The zeroth order term affects the vacuum fermion loop. The term linear in $a$ contributes to the Majoron one-point function; at the physical ground state the one-point function does not contribute to $V(\theta)$ by proper $\theta_0$ alignment. The quadratic term in $a$ gives a local $a^2\chi\chi$ interaction and provides the leading one-$M$ contribution to the zero-momentum Majoron two-point function. Further down, the expansion involves attaching more Majoron fields to the fermion, all at only one order of the explicit-breaking factor $M e^{-i\theta_0}$. In the small $M$ limit, this expansion provides multiple ways to generate an $n$-point one-loop function (see Fig.~\ref{fig:goldstone_diagrams}), among which amplitudes with a lower power of $M$ can be significant.

In Fig.~\ref{fig:goldstone_diagrams}, each diagram's number of vertices equals the power of $M$ in its amplitude. In the $M\ll m_0$ limit, the leading contribution to the $n$-point function is from the one with the fewest vertices, and significant interference also occurs between diagrams with equally few vertices. Naive vertex and propagator counting yields a low-momentum loop integrand $\propto v^{-n}(M/m_0)^m$ for an $n$-point diagram with $m$ vertices, which implies smooth symmetry restoration in the $M/m_0\rightarrow 0$ limit if the loop integrals do not introduce sensitivity to $M$.
The structure of the broken-phase propagator is derived from Eq.~\eqref{eq:Leff_theta}. The FNV propagators are
\begin{equation}
\langle \chi_\alpha(q)\chi_\beta(-q)\rangle
=
\frac{i m_0\,\epsilon_{\alpha\beta}}
{q^2-m_0^2+i\epsilon},
\qquad
\langle\chi^\dagger_{\dot\alpha}(q)
\chi^\dagger_{\dot\beta}(-q)\rangle
=
\frac{i m_0\,\epsilon_{\dot\alpha\dot\beta}}
{q^2-m_0^2+i\epsilon};
\label{eq:FNVbroken}
\end{equation}
fermion-number-preserving propagators are likewise determined by $m_0$~\cite{Dreiner:2008tw}. In this case, FNV refers to the propagator structure that changes chirality rather than the explicit symmetry breaking source.
The presence of a large $m_0$ is the major difference from Eq.~\ref{eq:FNV} in the previous section. $m_0$ protects the infrared safety of these loop integrals at $k=0$ and removes the type of small $M$ sensitivity as that in Eq. ~\ref{eq:zero_amp}. We will write out a few amplitudes to see it more explicitly.

\bigskip

For the Majoron mass, the 2-point contributions arise from the first two diagrams in Fig.~\ref{fig:goldstone_diagrams}, with couplings given by  
\begin{equation}
    \mathcal{L}_{aa\chi\chi}
=
\frac{M e^{-i\theta_0}}{4v^2}a^2\chi\chi
+
\mathrm{h.c.} \ \ {\text{and}}\ \ 
\mathcal{L}_{a\chi\chi}
=
+\frac{iM e^{-i\theta_0}}{2v}a\chi\chi+\mathrm{h.c.}
\end{equation}
Fig.~\ref{fig:single_insertion} contains only one insertion of the explicit-breaking
parameter and one FNV propagator. Thus it gives the leading $O(M)$ contribution:
\bea
i\Pi^{(2)}_{a^2\chi\chi,-}(0)
&=&
\frac{i M e^{-i\theta_0} m_0^3}{16\pi^2v^2}
\left(
1-\ln\frac{m_0^2}{\mu^2}
\right),
\label{eq:amp_single_majoron} \\
{\rm where}\ \ A_0(m_0^2)
&=&
\int\frac{d^dq}{(2\pi)^d}
\frac{1}{q^2-m_0^2+i\epsilon}
=
\frac{i\,m_0^2}{16\pi^2}
\left(
1-\ln\frac{m_0^2}{\mu^2}
\right),
\eea
by dimensional regularisation and \(\overline{\rm MS}\) subtraction.
The leading one-$M$ correction to the renormalized zero-momentum Majoron two-point function is
\begin{equation}
i\Pi^{(2)}_{a^2\chi\chi,-}(0)
=
\frac{iM e^{-i\theta_0} m_0^3}{16\pi^2v^2}
\left(
1-\ln\frac{m_0^2}{\mu^2}
\right).
\end{equation}
The $\theta_0$ dependence can be fixed by the vacuum alignment, to which we will return later. The explicit-breaking insertion supplies the single power of $M$. The remaining fermion-number-violating propagator is generated by the spontaneous mass $m_0$, giving the characteristic scaling $\Pi^{(1)}\propto M m_0^3/v^2$.
The next contribution from the bubble diagram (Fig.~\ref{fig:bubble}) contains two $a\chi\chi$ vertices:
\begin{equation}
i\Pi^{(2)}_{(1,1),-}(0)
=
\frac{1}{2!}\,
V_1^2\,
(-1)
\int\frac{d^dq}{(2\pi)^d}
\frac{
(im_0)^2
(\epsilon^{\alpha\beta}
\epsilon^{\gamma\delta}
\epsilon_{\beta\gamma}
\epsilon_{\delta\alpha})
}
{(q^2-m_0^2+i\epsilon)^2},
\end{equation}
with \(V_1=-Me^{-i\theta_0}/v\) read from
Eq.~\eqref{eq:vertex_tower_local} and \((-1)\) the closed-loop sign.
The spinor contraction gives
\(\epsilon^{\alpha\beta}\epsilon^{\gamma\delta}
\epsilon_{\beta\gamma}\epsilon_{\delta\alpha}=2\), and hence
\begin{equation}
i\Pi^{(2)}_{(1,1),-}(0)
=
\frac{M^2 e^{-2i\theta_0} m_0^2}{v^2}
\int\frac{d^dq}{(2\pi)^d}
\frac{1}{(q^2-m_0^2+i\epsilon)^2}.
\end{equation}
Using
\[
\int\frac{d^dq}{(2\pi)^d}
\frac{1}{(q^2-m_0^2+i\epsilon)^2}
=
-\frac{i}{16\pi^2}
\ln\frac{m_0^2}{\mu^2},
\]
the \((1,1)\) bubble contribution to the two-point function is at
$M^2$ order, for one orientation,
\begin{equation}
\Pi^{(2)}_{(1,1),-}(0)
=
-\frac{M^2 m_0^2}{16\pi^2v^2}
e^{-2i\theta_0}
\ln\frac{m_0^2}{\mu^2}.
\label{eq:amp_bubble_majoron}
\end{equation}
Adding the Hermitian-conjugate orientation,
\begin{equation}
\Pi^{(2)}_{(1,1)}(0)
=
-\frac{M^2 m_0^2}{16\pi^2v^2}
\ln\frac{m_0^2}{\mu^2}
\left(e^{-2i\theta_0}+e^{+2i\theta_0}\right).
\end{equation}
This visible bubble is not the complete fixed-order contribution.
With propagators defined using \(m_0\) alone, the same order also
contains the \((2,0)\) class, in which an \(a^2\chi\chi\) vertex
is accompanied by a zero-leg explicit-breaking insertion. Its
contribution is equal to that of the \((1,1)\) class,
\begin{equation}
\Pi^{(2)}_{(2,0)}(0)
=
-\frac{M^2 m_0^2}{16\pi^2v^2}
\ln\frac{m_0^2}{\mu^2}
\left(e^{-2i\theta_0}+e^{+2i\theta_0}\right).
\end{equation}
Similarly, other amplitudes with a single-propagator loop (Fig.~\ref{fig:3-0},~\ref{fig:4-0}) or a bubble loop (Fig.~\ref{fig:2-1},~\ref{fig:3-1},~\ref{fig:2-2}) topologies can be obtained by replacing the relevant multi-$a$-$\chi\chi$ coupling in the $e^{ia/v}$ expansion, which appears as fore-factors in Eq.~\ref{eq:amp_single_majoron} and ~\ref{eq:amp_bubble_majoron}, while the loop integral remains unchanged. The lowest order contribution to 3- and 4-point functions are also linear in $M$: 
\begin{align}
\Pi^{(3)}_{{\rm LO},-}(0)
&=
-\frac{iMe^{-i\theta_0} m_0^3}{16\pi^2v^3}
\left(1-\ln\frac{m_0^2}{\mu^2}\right),
\\[2mm]
\Pi^{(4)}_{{\rm LO},-}(0)
&=
-\frac{Me^{-i\theta_0} m_0^3}{16\pi^2v^4}
\left(1-\ln\frac{m_0^2}{\mu^2}\right),
\label{eq:Pi4_tadpole}
\end{align}
corresponding to Fig.~\ref{fig:3-0} and~\ref{fig:4-0},
respectively. The subscript `$_{-}$' denotes for the amplitude from one coupling term, without summing with that from its conjugate. These expressions denote a single orientation of the
explicit-breaking vertex; the Hermitian-conjugate sector must be
added to obtain the full real zero-momentum vertex. In particular,
\begin{equation}
\Pi^{(4)}_{\rm LO}(0)
=
-\frac{Mm_0^3}{16\pi^2v^4}
\left(1-\ln\frac{m_0^2}{\mu^2}\right)
\left(e^{-i\theta_0}+e^{+i\theta_0}\right),
\end{equation}
which at the aligned configuration \(\theta_0=\pi\) gives
\begin{equation}
\Pi^{(4)}_{\rm LO}(0)
=
\frac{Mm_0^3}{8\pi^2v^4}
\left(1-\ln\frac{m_0^2}{\mu^2}\right).
\end{equation}

\bigskip

The bubble integral still relies on a UV cut-off. The next higher order topology is the sub-sub-leading triangle diagram with three vertices. The contribution has the parametric structure
\begin{equation}
\Pi^{(3)}_{\rm triangle}(0)
=
-\,2\,\frac{M^3 m_0^3}{v^3}\,e^{-3i\theta_0}
\int\frac{d^dq}{(2\pi)^d}
\frac{1}{(q^2-m_0^2+i\epsilon)^3}
=
+\,i\,\frac{M^3 m_0}{16\pi^2 v^3}\,e^{-3i\theta_0}.
\label{eq:Pi3_triangle}
\end{equation}
Thus the triangle contributes at order \(M^3\).
For one cyclic ordering of the four external Majoron legs, the
same-orientation box contribution is
\begin{equation}
\Pi^{(4)}_{\rm box,-}\big|_{\rm 1\,cyc}(0)
=
+\,2i\,\frac{M^4 m_0^4}{v^4}\,e^{-4i\theta_0}
\int\frac{d^dq}{(2\pi)^d}
\frac{1}{(q^2-m_0^2+i\epsilon)^4}
=
-\,\frac{M^4}{48\pi^2 v^4}\,e^{-4i\theta_0}.
\label{eq:Pi4_box}
\end{equation}
Summing the three inequivalent cyclic orderings gives the complete
same-orientation box contribution,
\begin{equation}
\Pi^{(4)}_{\rm box,-}(0)
=
3\,\Pi^{(4)}_{\rm box,-}\big|_{\rm 1\,cyc}(0)
=
-\,\frac{M^4}{16\pi^2 v^4}\,e^{-4i\theta_0},
\end{equation}
as required by the \(q=4\) Coleman--Weinberg harmonic.
The infrared finiteness with $n>2$ is clearly provided by $m_0$ in the $M\ll m_0$ regime. To parameterise the scaling, the vertex weights $M/v^n$ for an $n$-Majoron
vertex and the loop integrals
$\int d^dq/(q^2-m_0^2)^n \sim 1/(16\pi^2 m_0^{2n-4})$ for $n\ge2$.
For the pure oriented \(n\)-polygon partition, in which each of the
\(q\) explicit-breaking vertices carries exactly one external
Majoron leg and the orientation is fixed, the amplitude satisfies
\begin{equation}
\Pi^{(n)}_{q,-}(0)
=
\mathcal O\!\left(
\frac{m_0^{4-q}}{16\pi^2}\left(\frac{-i}{v}\right)^n
M^q e^{-iq\theta_0}
\right),
\qquad q=1,2,\ldots,n,
\label{eq:Pi_power_counting_general}
\end{equation}
up to dimensionless numerical coefficients and logarithms of \(m_0^2/\mu^2\). 
At zero external momentum, the loop contribution to
the effective curvature is minus the loop self-energy,
$m_{a,\rm loop}^2=-\Pi_{\rm loop}^{(2)}$, where $\Pi_{\rm loop}^{(2)}$
is the sum of the two conjugate amplitudes
\begin{align}
m_{a,\rm loop}^2
={}&
-\frac{Mm_0^3}{16\pi^2v^2}
\left(1-\ln\frac{m_0^2}{\mu^2}\right)
\left(e^{-i\theta_0}+e^{+i\theta_0}\right)
\nonumber\\[2mm]
&+
\frac{M^2m_0^2}{8\pi^2v^2}
\ln\frac{m_0^2}{\mu^2}
\left(e^{-2i\theta_0}+e^{+2i\theta_0}\right)
+\mathcal O(M^3),
\label{eq:ma_final}
\end{align}
where \(m_{a,\rm loop}^2=-\Pi_{aa,\rm loop}(0)\). At
\(\theta_0=\pi\), this becomes
\begin{equation}
m_{a,\rm loop}^2
=
\frac{Mm_0^3}{8\pi^2v^2}
\left(1-\ln\frac{m_0^2}{\mu^2}\right)
+
\frac{M^2m_0^2}{4\pi^2v^2}
\ln\frac{m_0^2}{\mu^2}
+\mathcal O(M^3),
\end{equation}
in agreement with the corresponding expansion of the
Coleman--Weinberg curvature. Note that at the matching choice
\(\mu=m_0\), the \(M^2\) term vanishes identically; the factor-of-two
comparison is a general-\(\mu\) check.
In the $M \ll m_0$ limit, the leading Majoron mass squared is linear in the explicit-breaking parameter, $m_a^2\propto M$, and higher-order contributions (Fig.~\ref{fig:bubble}) are suppressed by powers of $M/m_0$.
While some finite coefficient in Eq.~\eqref{eq:ma_final} is scheme-dependent ($\overline{\rm MS}$), the $m_a^2 \propto M$ relation is scheme-independent, with the spontaneous scale $m_0=y v/\sqrt{2}$ preventing infrared sensitivity. Unlike Section~\ref{sec:model}, zero external momentum $k=0$ is convenient for higher $n$-point amplitudes. However, Eq.~\ref{eq:Pi4_tadpole},~\ref{eq:Pi_power_counting_general} do {\it not} mean all $n-$point amplitudes lead to a linear $\sim M \theta^n$ contribution to the effective potential near $\theta_0$. The amplitudes differ in complex phases and only their real part contribute, which relies on $\theta_0$. One can do so by restricting the summed one-point function to be zero. Or equivalently, by minimizing the $V(\theta)$ from path integral as we discuss below.

\bigskip

{\it Nonlinear realization comparison.} 
We can perform a quick match of these perturbative coefficients to path integral results by integrating out the fermions using the functional method
\cite{coleman1973radiative,Jackiw:1974cv}, i.e. the Coleman--Weinberg determinant. The zero-momentum projection of the one-particle-irreducible effective action determines the curvature of the effective potential
\cite{Iliopoulos:1974ur,Martin:2001vx,Masina:2025}. The determinant description provides a complementary viewpoint and makes the origin of the
fermion-number-violating (FNV) scaling transparent
\cite{Ellis:2020wxa}.
In the \(\overline{\rm MS}\) scheme, the Coleman--Weinberg contribution for a
single Majorana eigenstate whose field-dependent mass \(m(\theta)\)
is
\begin{equation}
V_{\rm CW}
=
-\frac{1}{32\pi^2}
m^4(\theta)
\left[
\ln\frac{m^2(\theta)}{\mu^2}
-\frac32
\right],
\qquad
m^2(\theta)=m_0^2+M^2+2m_0M\cos\theta
\label{eq:CW_Majorana}
\end{equation}
and $\theta$ is the relative phase. We first determine the minimum of this potential before performing the fluctuation expansion. The first derivative is
\begin{equation}
\frac{\partial V_{\rm CW}}{\partial\theta}
=
\frac{m_0M}{8\pi^2}
m^2(\theta)
\left[
\ln\frac{m^2(\theta)}{\mu^2}-1
\right]
\sin\theta .
\label{eq:CW_stationarity}
\end{equation}
The stationary points satisfy \(\sin\theta=0\) or
\(m^2(\theta)=e \mu^2\). The latter is almost always an unstable extremum due to
\begin{equation}
\left.
\frac{\partial^2V_{\rm CW}}{\partial\theta^2}
\right|_{m^2=\exp(1)\mu^2}
=
-\frac{m_0^2M^2}{4\pi^2}\sin^2\theta
\le 0 ,
\label{eq:CW_interior_max}
\end{equation}
and the minima
are restricted to \(\theta_s=0,\pi\) on the compact domain, which correspond to
$m_s^2=(m_0\pm M)^2,$
Stability still relies on the sign of $\ln{m^2/\mu^2}-1$. We can choose it to be negative for a large $\mu$ so that the large negative $-\ln\mu^2$ term dominates the bracket, and the physical minimum is at \(\theta_{\rm CW}=\pi\). We do not consider the small $\mu$ case as it introduces more physics below the symmetry breaking scale. Expanding \(\theta=\pi+\delta\theta\) gives
\begin{equation}
m^2(\pi+\delta\theta)
=
m_0^2+M^2-2m_0M\cos\delta\theta
=
(m_0-M)^2
+
m_0M\,\delta\theta^2
-
\frac{m_0M}{12}\delta\theta^4
+\mathcal O(\delta\theta^6).
\label{eq:m2_CW_min_expansion}
\end{equation}
so all odd powers in the Taylor expansion vanish. The perturbative
expansion of the potential \(V_{\rm CW}\) gives
\begin{align}
V_{\rm CW}(\pi+\delta\theta)
&=
V_0
+
\frac12 V_2\,\delta\theta^2
+
\frac{1}{24}V_4\,\delta\theta^4
+
\mathcal O(\delta\theta^6),
\label{eq:CW_expansion_minimum}
\intertext{where}
V_0
&=
-\frac{1}{32\pi^2}(m_0-M)^4
\left(\ln\frac{(m_0-M)^2}{\mu^2}-\frac32\right),
\label{eq:V_0}
\intertext{and}
V_2
&=
\frac{m_0M\,(m_0-M)^2}{8\pi^2}
\left(1-\ln\frac{(m_0-M)^2}{\mu^2}\right),
\label{eq:CW_V2_minimum}
\\
V_4
&=
-V_2
-
\frac{3m_0^2M^2}{4\pi^2}
\ln\frac{(m_0-M)^2}{\mu^2}.
\label{eq:CW_V4_minimum}
\end{align}
See Appendix~\ref{app:CW_compact} for details. The Majoron
potential's mass and quartic coefficients are
\begin{align}
m_{a,\rm CW}^2
&=
\frac{V_2}{v^2}
=
\frac{m_0M\,(m_0-M)^2}{8\pi^2v^2}
\left(1-\ln\frac{(m_0-M)^2}{\mu^2}\right),
\label{eq:ma_CW_exact_min}
\\
\lambda_{a,\rm CW}
&=
\frac{V_4}{24v^4}
=
-\frac{m_{a,\rm CW}^2}{24v^2}
-
\frac{m_0^2M^2}{32\pi^2v^4}
\ln\frac{(m_0-M)^2}{\mu^2}.
\label{eq:lambda_CW_exact_min}
\end{align}
Since $M\ll m_0$, one may further expand to higher powers of
\(M\),
\begin{equation}
m^2_{a,\rm CW}(\pi)
=
\frac{m_0^3M}{8\pi^2v^2}(1-L)
+
\frac{m_0^2M^2}{4\pi^2v^2}L
-
\frac{m_0M^3}{8\pi^2v^2}(L+2)
+
\frac{M^4}{12\pi^2v^2}
+
\mathcal O\!\left(\frac{M^5}{m_0v^2}\right),
\label{eq:ma_CW_power}
\end{equation}
where \(L\equiv\ln(m_0^2/\mu^2)\).

\bigskip

$V_{\rm CW}(\theta)$ expansion should match the 1-loop $n$-point Majoron functions at zero temperature. To see how it matches the explicit amplitude results, we can take $\theta_0\rightarrow \pi$ into Eq.~\ref{eq:amp_single_majoron}-\ref{eq:ma_final}. For an \(n\)-point diagram with \(q\) vertices, Eq.~\eqref{eq:Pi_power_counting_general} gives at
\(\theta_0=\pi\) the complex argument
\begin{equation}
\mathrm{Arg}\!\left(\Pi^{(n)}_{q,-}(0)\big|_{\theta_0=\pi}\right)
=
\left(\frac{n}{2}-q\right)\pi,
\qquad q=1,2,\ldots,n,
\end{equation}
up to the real numerical coefficient. For odd \(n\) and integer
\(q\), this is a half-integer multiple of \(\pi\) and makes the amplitude imaginary; the conjugate-summed
\(n\)-point function therefore vanishes for odd \(n\), in agreement
with the even structure of $V_{\rm CW}$ expansion.
To the leading order, the quartic coefficient at
\(\theta_{\rm CW}=\pi\) is
\begin{equation}
\lambda_{a,\rm CW}
=
-\frac{Mm_0^3}{192\pi^2v^4}
\left(1-\ln\frac{m_0^2}{\mu^2}\right)+{\cal O}(M^2).
\end{equation}
Since \(\lambda_{a,\rm CW}\) multiplies \(a^4\) in the potential,
it should not be compared directly with the zero-momentum four-point
1PI vertex. Four differentiation's give a factor of \(4!\), and with
the 1PI sign convention used above,
\begin{equation}
\Pi^{(4)}_{\rm LO}(0)
=
-\,4!\,\lambda_{a,\rm CW}^{(1)}
=
\frac{Mm_0^3}{8\pi^2v^4}
\left(1-\ln\frac{m_0^2}{\mu^2}\right),
\end{equation}
which agrees with the diagrammatic result. 

\section{Conclusion}
\label{sec:conclusion}

To summarize, this work provides diagrammatic one loop level calculations for the restoration of the combined phase-shift symmetry in the light Majorana fermion limit. As the explicit symmetry breaking parameter, a nonzero yet small Majorana mass allows the vacuum Majorana fermion loop to develop infrared sensitivities that contribute to the effective symmetry potentials when the Majorana fermion couples to other fields. We consider two typical scenarios for a Majorana fermion coupled to a scalar: (i) a perturbative scalar scenario with a point minimum potential; (ii) a spontaneous symmetry breaking scenario with the Nambu-Goldstone mode as the Majoron. We perform explicit Feynman diagram calculations at the one loop level, and demonstrate the continuous vanishing of the symmetry breaking amplitudes with in the soft symmetry breaking parameter limit $M\rightarrow 0$, which agrees with 't Hooft naturalness.

The zero temperature radiative Majorana fermion loop faces an infrared sensitivity issue when the Majorana fermion is taken to be light, and it worsens with more fermion number violating propagators inside the loop. Without a regulatory kinematic scale, the sensitivity lies with the symmetry breaking parameter $M$, causing large symmetry breaking amplitudes to persist regardless of the smallness of $M$. In the coupled theory, the naturally existing scales, e.g. scalar mass near the potential minimum, can conveniently serve as a reference kinematic point to match the symmetry breaking amplitudes, in which we demonstrate that such Majorana fermion one-loop amplitudes are infrared finite, and that they scale with a positive power of $M/m_\phi$, continuously restoring the joint shift symmetry in the small $M$ limit.

For calculations, we organized the symmetry breaking diagrams by their $(N_{\rm FNV},N_{\rm FNP})$ propagator structure, which provides a transparent framework to iterate the contributing amplitude to the powers of the explicit symmetry breaking parameter.
The radiative Majoron potential exhibits a simple organizing principle: each order in the explicit-breaking expansion is associated with a definite number of explicit-breaking insertions.  The overall scaling is inherited from the corresponding FNV structure. The single-insertion, pure bubble, pure triangle, and pure FNV box diagrams generate the leading contributions at orders $M$, $M^2$, $M^3$, $M^4$ and so forth. We calculate the amplitudes explicitly at fixed nonzero external momenta up to the $n=6$ hexagon diagram, and show that they align with the general scaling relation.

In the spontaneous symmetry breaking scenario, the radial scalar VEV makes the chiral-rotated fermion field massive, which already prevents Majorana loop amplitudes from infrared divergences. The corresponding Majoron $n$-point functions are protected from infrared divergences by the scalar VEV and demonstrate positive power counting with each explicit symmetry breaking vertex. Thus symmetry is naturally restored in the $M/v\rightarrow 0$ limit. The effective Majoron potential, along with each $n$-point Majoron function that breaks the shift symmetry, demonstrates linear ${\cal O}(M/v)$
dependence at the leading order, which agrees with the perturbative expansion of a nonlinear Coleman-Weinberg determinant near the $V(\theta)$ minimum.

Viewed from this perspective, the naturally existing scales of the scalar sector, which corresponds to the shape of scalar potential --- the mass or the VEV, help regulating the infrared of physical symmetry breaking amplitudes that contain Majorana fermion loop. 
A completely flattened scalar potential would lack such a regulatory scale, where the Majorana mass $M$ appears as the only scale of the theory that always violates the phase shift symmetry. 

Future work can extend this framework to include finite-temperature effects, the dynamics of the radial mode, and a detailed exploration of the parameter space that resolves dark matter and other phenomenology.

\begin{acknowledgments}
This work is supported by the National Natural Science Foundation of China under Grant No.~12447105.
\end{acknowledgments}

\bibliographystyle{apsrev4-2}
\bibliography{bib}
\appendix

\setcounter{equation}{0}
\renewcommand{\theequation}{A\arabic{equation}}

\section{One-loop matching functions in the zero-VEV regime}
\label{app:loops}

The one-loop bubble, triangle, and box functions that are pertinent to symmetry-violating scalar operators in the zero-VEV domain are presented here. We estimate the associated 1PI vertex functions at the given on-shell, or analytically continued, kinematics $p_i^2=m_\phi^2$, assuming the hierarchy $m_\phi\gg M$. The $\overline{\rm MS}$ method and $d=4-2\varepsilon$ are utilized for dimensional regularization throughout~\cite{tHooft1972dimreg,Bollini1972dimreg}.
The main text's Eqs.~\eqref{eq:FNP} and \eqref{eq:FNV} provide the fermion propagators.

\subsection{General loop formalism and power counting}
\label{app:general_feynman}
Scalar loop integrals are the reduction of all pure-FNV diagrams. The $n$-point scalar integral with propagators uses the conventional one-loop form. All pure-FNV diagrams reduce to scalar loop integrals after the trace of simple epsilon that gives 2. By following the standard one-loop formalism~\cite{Passarino:1978jh,tHooft:1978jhc,Denner1993}, the $n$-point scalar integral with propagator's denominator $D_j=(q+k_j)^2-M^2+i\epsilon$ is
\begin{equation}
I_n =
\int\!\frac{d^dq}{(2\pi)^d}\,
\prod_{j=1}^{n}\frac{1}{D_j},
\label{eq:general_In}
\end{equation}
with $k_1=0$ and the external momenta passing through $k_2,...,k_n$. Feynman parameters $x_i\ge0$ with $\sum_i x_i=1$ are introduced. Using the identity
\begin{equation}
\frac{1}{\prod_{j=1}^n D_j}
=
(n-1)!\int_0^1[dx]\,
\frac{1}{\bigl(\sum_{j=1}^n x_j D_j\bigr)^n},
\qquad
[dx]
=
dx_1\cdots dx_n\,
\delta\!\left(1-\sum_{i=1}^n x_i\right).
\end{equation}
After shifting the loop momentum $q\to q'=q+\sum_jx_jk_j$, the denominator becomes $(q')^2-\Delta+i\epsilon$, with
\begin{equation}
\Delta
=
M^2-\sum_{i<j}x_ix_j(k_i-k_j)^2 .
\label{eq:Delta_general}
\end{equation}

In $d=4-2\varepsilon$ dimensions, the standard momentum integral gives
\begin{equation}
\int\!\frac{d^dq'}{(2\pi)^d}\,
\frac{1}{[(q')^2-\Delta+i\epsilon]^n}
=
\frac{i(-1)^n}{(4\pi)^2}\,
\frac{\Gamma(n-2)}{\Gamma(n)}\,
\frac{1}{(\Delta-i\epsilon)^{n-2}}
\left[1+\mathcal O(\varepsilon)\right].
\end{equation}
Thus, for $n\ge3$, the scalar integrals may be written schematically as
\begin{equation}
I_n
=
\frac{i}{16\pi^2}\,
\frac{\Gamma(n-2)}{m_\phi^{2n-4}}
\int d\Pi_n\,
\frac{1}{\bigl(f_n(x)+r\bigr)^{\,n-2}},
\qquad
r=\frac{M^2}{m_\phi^2}\ll1,
\label{eq:In_scaled}
\end{equation}
up to the overall sign that the selected analytical continuation suggests. In this case, $d\Pi_n\equiv[dx]$ and $f_n(x)$ represent the dimensionless kinematic function suitable for the selected momentum configuration. While the equivalent $i\epsilon$ prescription is understood for physical timelike kinematics, we select the convention $f_n(x)\ge0$ for the Euclidean symmetric configuration employed below, resulting in the denominator taking the form $r+f_n(x)$. The $n=2$ scenario is handled differently and is logarithmically ultraviolet divergent.

The superficial degree of divergence is
\begin{equation}
D=4+N_q-2n,
\end{equation}
where the power of loop momentum in the numerator is denoted by $N_q$. Pure-FNV diagrams with $n\ge3$ are UV finite since $N_q=0$, while the Feynman-parameter endpoint structure governs their small-$M$ behavior. Specifically, the parameter measure and the zero of $f_n(x)$ must be included close to an isolated simplex vertex, which results in a finite $F_3(0)$ with subleading $\mathcal O(r\ln r)$ corrections for the pure triangle and a finite $F_4(0)$ with $\mathcal O(r\ln r)$ corrections for the pure box. The resulting behavior is summarized in Table~\ref{tab:log_summary}.

\begin{table}[h]
\centering
\caption{UV behaviour and explicit-breaking power counting of
the leading zero-VEV loop functions.}
\small
\begin{tabular}{lcccc}
\toprule
Diagram & $D$ & UV behaviour & Small-$r$ structure & Leading scaling \\
\midrule
Bubble & $0$ & Log.\ divergent &
$\ln(m_\phi^2/\mu^2)$ & $M^2$ \\
Mixed triangle & $0$ & Log.\ divergent &
$\ln(m_\phi^2/\mu^2)$ & $M$ \\
Pure triangle & $-2$ & UV finite &
$\ln(m_\phi^2/M^2)$ & $M^3/m_\phi^2$ \\
Pure box & $-4$ & UV finite &
$F_4(0)+\mathcal O(r\ln r)$ & $M^4/m_\phi^4$ \\
\bottomrule
\end{tabular}
\label{tab:log_summary}
\end{table}

Let $N_M$ be the number of FNV mass factors for an $n$-point scalar operator produced by a fermion loop, and let $\mathcal N_n(x)$ be any numerator structure resulting from the spinor trace. The matching coefficient scales as
\begin{equation}
\lambda_n
\sim
\frac{y^n M^{N_M}}
{16\pi^2\,m_\phi^{2n-4}}
\int d\Pi_n\,
\frac{\mathcal N_n(x)}
{\bigl(f_n(x)+r\bigr)^{\,n-2}},
\qquad
r=\frac{M^2}{m_\phi^2}.
\label{eq:master_formula}
\end{equation}

\subsubsection{Bubble diagram}

The bubble comprises two external $\phi$ legs and two FNV propagators having scalar integral as
\begin{equation}
I_2(p^2)
=
\int \frac{d^d q}{(2\pi)^d}
\frac{1}{D_1D_2},
\qquad
D_1=q^2-M^2+i\epsilon,
\quad
D_2=(q+p)^2-M^2+i\epsilon,
\end{equation}
with $p^2=m_\phi^2$. Using the Feynman parametrisation
\begin{equation}
\frac{1}{D_1D_2}
=
\int_0^1 dx\,
\frac{1}{[xD_1+(1-x)D_2]^2},
\end{equation}
we obtain
\begin{equation}
xD_1+(1-x)D_2
=
(q+(1-x)p)^2-\Delta,
\qquad
\Delta=M^2-x(1-x)p^2.
\end{equation}
Shifting the loop momentum $q\to\ell=q+(1-x)p$ yields
\begin{equation}
I_2(p^2)
=
\int_0^1 dx
\int\frac{d^d\ell}{(2\pi)^d}
\frac{1}{(\ell^2-\Delta+i\epsilon)^2}.
\end{equation}

The momentum integral is standard~\cite{tHooft:1978jhc}:
\begin{equation}
\int \frac{d^d\ell}{(2\pi)^d}
\frac{1}{(\ell^2-\Delta+i\epsilon)^2}
=
\frac{i}{16\pi^2}
\left[
\frac{1}{\varepsilon}
-\gamma_E
+\ln(4\pi)
-\ln\frac{\Delta-i\epsilon}{\mu^2}
\right]
+\mathcal O(\varepsilon).
\end{equation}
After $\overline{\rm MS}$ subtraction,
\begin{equation}
I_2^{\rm ren}(p^2)
=
-\frac{i}{16\pi^2}
\int_0^1 dx\,
\ln\frac{\Delta-i\epsilon}{\mu^2}.
\end{equation}
In the hierarchy $m_\phi^2\gg M^2$,
\begin{equation}
\Delta
\simeq
-x(1-x)m_\phi^2.
\end{equation}
The Feynman-parameter integral for the
\begin{equation}
\int_0^1 dx\,\ln[x(1-x)]
=
-2,
\label{eq:bubble_log_integral}
\end{equation}
so that
\begin{equation}
I_2^{\rm ren}(m_\phi^2)\big|_{\rm disp}
=
\frac{i}{16\pi^2}
\left[
-\ln\frac{m_\phi^2}{\mu^2}
+2
\right].
\end{equation}

\subsubsection{Pure triangle diagram}

The pure triangle $(N_{\rm FNV},N_{\rm FNP})=(3,0)$ contains three
FNV propagators with the external momentum configuration
\[
p_1=-p_2\equiv p,\qquad p_3=0,
\qquad p^2=m_\phi^2,
\]
the amplitude can therefore be written as
\begin{equation}
i\mathcal M_3(p)
=
y^3M^3\,I_3(p^2),
\label{eq:pure_triangle_amp}
\end{equation}
where
\begin{equation}
I_3(p^2)
=
\int\frac{d^dq}{(2\pi)^d}
\frac{1}
{
(q^2-M^2+i\epsilon)^2
[(q+p)^2-M^2+i\epsilon]
}.
\label{eq:pure_triangle_scalar}
\end{equation}

The superficial degree of divergence is
\[
D_{\rm UV}=4-2(3)=-2,
\]
so this scalar triangle integral is ultraviolet finite. Applying Feynman parametrisation,
\begin{equation}
\frac{1}{D_1^2D_2}
=
2\int_0^1 dx\,
\frac{x}{[xD_1+(1-x)D_2]^3},
\end{equation}
with
\[
D_1=q^2-M^2+i\epsilon,
\qquad
D_2=(q+p)^2-M^2+i\epsilon,
\]
we have
\begin{equation}
xD_1+(1-x)D_2
=
\bigl[q+(1-x)p\bigr]^2
-
\Delta
+i\epsilon,
\end{equation}
where
\begin{equation}
\Delta
=
M^2-x(1-x)p^2 .
\end{equation}

After shifting
\[
\ell=q+(1-x)p,
\]
the integral becomes
\begin{equation}
I_3(p^2)
=
2\int_0^1 dx\,x
\int\frac{d^d\ell}{(2\pi)^d}
\frac{1}{(\ell^2-\Delta+i\epsilon)^3}.
\end{equation}
In four dimensions,
\begin{equation}
\int\frac{d^4\ell}{(2\pi)^4}
\frac{1}{(\ell^2-\Delta+i\epsilon)^3}
=
-\frac{i}{32\pi^2}
\frac{1}{\Delta-i\epsilon},
\end{equation}
and hence
\begin{equation}
I_3(p^2)
=
-\frac{i}{16\pi^2}
\int_0^1 dx\,
\frac{x}
{M^2-x(1-x)p^2-i\epsilon}.
\label{eq:pure_tri_integral}
\end{equation}

For the kinematic point $p^2=m_\phi^2$, define
\[
r=\frac{M^2}{m_\phi^2},
\qquad
\beta=\sqrt{1-4r}.
\]
When $m_\phi^2>4M^2$, the integral can be evaluated as
\begin{equation}
I_3(m_\phi^2)
=
-\frac{i}{16\pi^2m_\phi^2\beta}
\left[
\ln\frac{1-\beta}{1+\beta}
+i\pi
\right].
\label{eq:pure_triangle_exact}
\end{equation}
The $i\pi$ term arises from the open two-fermion cut in the timelike
$p^2$ channel. With the hierarchy $M^2\ll m_\phi^2$,
\[
\beta=1+\mathcal O\!\left(\frac{M^2}{m_\phi^2}\right),
\qquad
\ln\frac{1-\beta}{1+\beta}
=
\ln\frac{M^2}{m_\phi^2}
+
\mathcal O\!\left(\frac{M^2}{m_\phi^2}\right),
\]
so that
\begin{equation}
I_3(m_\phi^2)
=
-\frac{i}{16\pi^2m_\phi^2}
\left[
\ln\frac{M^2}{m_\phi^2}
+i\pi
\right]
+
\mathcal O\!\left[
\frac{iM^2}{16\pi^2m_\phi^4}
\ln\frac{m_\phi^2}{M^2}
\right].
\label{eq:pure_triangle_smallM}
\end{equation}
Equivalently, with the amplitude convention of
Eq.~\eqref{eq:pure_triangle_amp},
\begin{equation}
\mathcal M_3
=
\frac{y^3M^3}{16\pi^2m_\phi^2}
\left[
\ln\frac{m_\phi^2}{M^2}
-i\pi
+
\mathcal O\!\left(
\frac{M^2}{m_\phi^2}
\ln\frac{m_\phi^2}{M^2}
\right)
\right].
\label{eq:pure_triangle_final}
\end{equation}

\subsubsection{Mixed triangle}

The mixed triangle $(N_{\rm FNV},N_{\rm FNP})=(1,2)$ diagram contains
one FNV propagator and two FNP propagators. We use the kinematic
configuration
\[
p_1=p,\qquad p_2=-p,\qquad p_3=0,
\qquad p^2=m_\phi^2,
\]
so that the two nonzero external scalar legs satisfy
$p_1^2=p_2^2=m_\phi^2$, while the third leg is a zero-momentum off-shell
insertion. Choosing the loop routing such that the two FNP propagators carry
momenta $q$ and $q+p$, while the FNV propagator carries $q$, the
spinor trace gives
\[
\operatorname{Tr}\!\left[
\sigma\!\cdot q\,\bar\sigma\!\cdot(q+p)
\right]
=
2q\!\cdot(q+p).
\]
The amplitude can therefore be written as
\begin{equation}
i\mathcal M_{3}^{\rm mix}
=
y^3M\, I_{\rm mix}(p^2),
\end{equation}
with
\begin{equation}
 I_{\rm mix}(p^2)
=
\mu^{2\epsilon}
\int\frac{d^dq}{(2\pi)^d}
\frac{2q\!\cdot(q+p)}
{(q^2-M^2+i\epsilon)^2
[(q+p)^2-M^2+i\epsilon]}.
\label{eq:mixed_triangle_amp}
\end{equation}
Using
\[
2q\!\cdot(q+p)
=
(q^2-M^2)
+
[(q+p)^2-M^2]
+
2M^2-p^2,
\]
the ultraviolet divergent part is
\[
I_{\rm mix}^{\rm div}
=
\frac{i}{8\pi^2}\frac{1}{\bar\epsilon},
\]
showing explicitly that the mixed triangle is logarithmically
ultraviolet divergent. After $\overline{\rm MS}$ subtraction and setting $p^2=m_\phi^2$, define
\[
r=\frac{M^2}{m_\phi^2},
\qquad
\beta=\sqrt{1-4r}.
\]
For $m_\phi^2>4M^2$, the renormalized loop contribution is
\begin{equation}
 I_{\rm mix}^{\overline{\rm MS}}(m_\phi^2)
=
\frac{i}{8\pi^2}
\left[
1-\ln\frac{M^2}{\mu^2}
-
\frac{1-3r}{\beta}
\left(
\ln\frac{1+\beta}{1-\beta}
-i\pi
\right)
\right].
\label{eq:mixed_triangle_exact}
\end{equation}
In the symmetry-restoring regime $M^2\ll m_\phi^2$, using
\[
\ln\frac{1+\beta}{1-\beta}
=
\ln\frac{m_\phi^2}{M^2}
+\mathcal O(r),
\]
the logarithms of $M$ cancel, giving
\begin{equation}
I_{\rm mix}^{\overline{\rm MS}}(m_\phi^2)
=
\frac{i}{8\pi^2}
\left[
1-\ln\frac{m_\phi^2}{\mu^2}
+i\pi
+
\mathcal O\!\left(
\frac{M^2}{m_\phi^2}
\ln\frac{M^2}{m_\phi^2}
\right)
\right].
\label{eq:mixed_triangle_smallM}
\end{equation}
Consequently, the mixed triangle scales linearly with the explicit mass M:
\begin{equation}
    \mathcal M_3^{\rm mix}
=
\mathcal O(y^3M),
\end{equation}

\subsection{Four-point functions: the box diagram}
\label{app:box}

\subsubsection{Adjacent mixed box}

The adjacent mixed box has
\[
(N_{\rm FNV},N_{\rm FNP})=(2,2),
\]
with the fermion loop having the two adjacent FNV propagators. The symmetry-violating structure produced is
\[
\phi^3\phi^*+\mathrm{h.c.}
\]
then the momentum routing
\[
k_1=0,\qquad k_2=p_1,\qquad k_3=p_1+p_2,\qquad k_4=-p_4,
\]
with $p_i^2=m_\phi^2$, the amplitude is
\begin{equation}
i\mathcal{M}_{\rm adj}
\sim
y^4M^2
\int\!\frac{d^dq}{(2\pi)^d}\,
\frac{N_{\rm adj}(q,k_i)}
{\prod_{j=1}^{4}D_j},
\label{eq:adj_box_amp}
\end{equation}
where $D_j=(q+k_j)^2-M^2+i\epsilon$. The two FNV propagators give $M^2$, while the two FNP propagators yield $N_{\rm adj}\sim q^2$ at loop momentum. Thus
\begin{equation}
D=4+2-2(4)=-2,
\end{equation}
so the integral is ultraviolet finite. Following tensor reduction and Feynman parametrization, the contribution is
\begin{equation}
i\mathcal{M}_{\rm adj}
=
\frac{y^4}{8\pi^2}
\frac{M^2}{m_\phi^2}
F_{4,\rm adj}\!
\left(\frac{M^2}{m_\phi^2}\right),
\label{eq:F4_adj}
\end{equation}
where
\begin{equation}
F_{4,\rm adj}(r)
=
F_{4,\rm adj}(0)
+\mathcal O(r\ln r),
\qquad
r=\frac{M^2}{m_\phi^2}.
\end{equation}

\subsubsection{Opposite mixed box}

The opposite mixed box has
\[
(N_{\rm FNV},N_{\rm FNP})=(2,2),
\]
However, FNP propagators surround the loop, separating the two FNV propagators. The external-field contraction is altered by this new ordering:
The phase-invariant structure is influenced by the opposing topology.
\[
\phi^2\phi^{*2}=|\phi|^4,
\]
rather than $\phi^3\phi^*+\mathrm{h.c.}$
Its amplitude is
\begin{equation}
i\mathcal{M}_{\rm opp}
\sim
y^4M^2
\int\!\frac{d^dq}{(2\pi)^d}\,
\frac{N_{\rm opp}(q,k_i)}
{\prod_{j=1}^{4}D_j},
\label{eq:opp_box_amp}
\end{equation}
with $N_{\rm opp}\sim q^2$ at large momentum, so $D=-2$. The coefficient is
\begin{equation}
i\mathcal{M}_{\rm opp}
=
\frac{y^4}{8\pi^2}
\frac{M^2}{m_\phi^2}
F_{4,\rm opp}\!
\left(\frac{M^2}{m_\phi^2}\right),
\label{eq:F4_opp}
\end{equation}
where
\begin{equation}
F_{4,\rm opp}(r)
=
F_{4,\rm opp}(0)
+\mathcal O(r\ln r).
\end{equation}
This topology remains finite as $M\to0$, but since it multiplies the
invariant operator $|\phi|^4$, it does not generate a phase-dependent
symmetry-violating interaction.

\subsubsection{Pure-FNV box}

The pure box $(4,0)$ generates $\phi^4+\phi^{*4}$. The external legs
satisfy $p_i^2=m_\phi^2$ and $p_1+p_2+p_3+p_4=0$. The amplitude is
\begin{equation}
i\mathcal M_4^{\rm box}
=
y^4M^4\,I_4,
\qquad
I_4
=
\int\frac{d^dq}{(2\pi)^d}
\frac{1}{D_1D_2D_3D_4}.
\end{equation}
The scalar integral has
$D=4-2\times4=-4$ and is ultraviolet finite. Feynman parametrisation
gives
\begin{equation}
I_4
=
\frac{i}{16\pi^2}
\int d\Pi_4\,
\frac{1}{\Delta^2},
\qquad
\Delta=m_\phi^2\bigl(r+f_4(x_i)\bigr),
\end{equation}
with
\begin{equation}
F_4(r)
=
\int d\Pi_4\,
\frac{1}{\bigl(r+f_4(x_i)\bigr)^2}.
\label{eq:F4_def_app}
\end{equation}
Thus
\begin{equation}
I_4
=
\frac{i}{16\pi^2m_\phi^4}
F_4\!\left(\frac{M^2}{m_\phi^2}\right).
\end{equation}

\subsubsection{Numerical limits of box topologies}

The placement of the FNV propagators distinguishes the three box
topologies. For the symmetric four-point configuration
\[
p_i^2=m_\phi^2,\qquad p_i\cdot p_j=-\frac{m_\phi^2}{3},
\]
the box functions are
\begin{equation}
F_{4,a}(r)
=
\int d\Pi_4
\frac{N_a(x_i)}
{\bigl(r+f_4(x_i)\bigr)^2},
\qquad
a\in\{\mathrm{pure},\mathrm{adj},\mathrm{opp}\}.
\end{equation}
The reduced numerators for the mixed boxes are
\begin{equation}
N_{\rm adj}
=
1-2x_1x_3-2x_2x_4,
\qquad
N_{\rm opp}
=
1-2x_1x_2-2x_3x_4.
\end{equation}
Deterministic adaptive integration over the simplex gives
\begin{equation}
F_4^{\rm pure}(0)=1.91,\ \ 
F_{4,\rm adj}(0)=1.61,\ \ \text{and~} 
F_{4,\rm opp}(0)=1.59 .    
\end{equation}
The finite limit
\begin{equation}
F_{4,a}(r\rightarrow0)=F_{4,a}(0)
\end{equation}
shows that the box functions contain no inverse powers of
$r=M^2/m_\phi^2$. Such terms would generate infrared enhancements
capable of compensating the explicit powers of $M$ carried by the FNV
propagators~\cite{Smirnov:2002pj,Friot:2005cu}. Instead, the leading
non-analytic correction is of order
$r\ln r$,
\begin{equation}
F_{4,a}(r)
=
F_{4,a}(0)+\mathcal O(r\ln r),
\end{equation}
so $F_{4,a}(r)=\mathcal O(1)$ in the symmetry-restoring limit.

\subsection{Small-$r$ expansion of the pure box function}
\label{app:box_expansion}

For the symmetric configuration, the Symanzik polynomial is
\begin{equation}
f_4(x_1,x_2,x_3,x_4)
=
x_1x_2+\frac43 x_1x_3+x_1x_4+x_2x_3+\frac43 x_2x_4+x_3x_4,
\label{eq:f4_symanzik}
\end{equation}
with $x_1+x_2+x_3+x_4=1$ and $x_i\ge0$. Since all coefficients are
positive, $f_4=0$ only at the four vertices. Near $x_1=1$, parametrise
\[
x_1=1-t,\qquad x_2=tu,\qquad x_3=tv,\qquad x_4=t(1-u-v).
\]
The measure becomes $d\Pi_4=t^2\,dt\,du\,dv$, and expanding $f_4$ gives
\[
f_4=t\,g(u,v)+\mathcal O(t^2),\qquad
g(u,v)=1+\frac13 v>0.
\]
The endpoint contribution is
\[
F_4^{(\mathrm{endpoint})}(r)
=
\int du\,dv
\int_0^1dt\,
\frac{t^2}{(r+t\,g)^2}.
\]
With $y=r+t g$, the $t$-integral becomes
\begin{equation}
\int_0^1dt\,
\frac{t^2}{(r+t\,g)^2}
=
\frac{1}{g^3}
\int_r^{r+g}dy\,
\frac{(y-r)^2}{y^2}.
\end{equation}
Using
\[
\frac{(y-r)^2}{y^2}
=
1-\frac{2r}{y}+\frac{r^2}{y^2},
\]
one obtains
\begin{equation}
\int_0^1dt\,
\frac{t^2}{(r+t\,g)^2}
=
\frac{1}{g^2}
+
\frac{2r}{g^3}\ln r
+
\mathcal O(r),
\label{eq:endpoint_integral}
\end{equation}
where the $r\ln g$ term is absorbed into $\mathcal O(r)$. The bulk region
is analytic in $r$, and the remaining vertices give identical endpoint
contributions. Hence
\begin{equation}
F_4(r)
=
F_4(0)
+
\mathcal O(r\ln r).
\label{eq:F4_expansion}
\end{equation}
No inverse powers of $r$ arise, so $\lambda_4\propto M^4$.
The coefficient of $r\ln r$ may be extracted via Mellin--Barnes or
sector-decomposition techniques
\cite{Smirnov:2002pj,Binoth:2000ps,Friot:2005cu}. Deterministic
adaptive integration gives
\begin{equation}
F_4(0)=1.91,
\label{eq:F4_zero}
\end{equation}
and fitting to
\begin{equation}
F_4(r)
=
1.91
+c_1\,r\ln r
+c_2\,r
+c_3\,r^2\ln r
+\mathcal O(r^2)
\label{eq:F4_fit_full}
\end{equation}
yields
\[
c_1=3.10,\qquad
c_2=-1.77,\qquad
c_3=-2.28.
\]
Varying the fitting window gives $c_1=3.18$, $3.09$, and $3.05$ for
$r<10^{-2}$, $r<5\times10^{-3}$, and $r<2\times10^{-3}$, respectively.
The essential result is the absence of inverse powers of $r$; the loop
function cannot compensate the explicit $M^4$ suppression.

\subsection{Pure-FNV pentagon and hexagon}
\label{app:higher_pure}

For completeness, we also consider the pure-FNV pentagon and hexagon
topologies. These contain five and six FNV propagators, respectively,
and therefore carry five and six explicit powers of the
fermion-number-violating Majorana mass $M$. The pentagon is evaluated
at a fixed physical $2\to3$ scattering configuration, while the hexagon
is evaluated at the Euclidean symmetric point, which is analytically
continued to the physical region for even-point functions. All external
scalar legs satisfy
\[
p_i^2=m_\phi^2,
\qquad
\sum_i p_i=0 .
\]
For the pure pentagon, $(N_{\rm FNV},N_{\rm FNP})=(5,0)$, the amplitude is
\begin{equation}
i\mathcal M_5^{\rm pent}
=
y^5M^5 I_5,
\qquad
I_5
=
\int\frac{d^dq}{(2\pi)^d}
\frac{1}{D_1D_2D_3D_4D_5},
\end{equation}
where
\[
D_j=(q+K_j)^2-M^2+i\epsilon .
\]
The superficial degree of divergence is
\[
D_{\rm UV}=4-2(5)=-6,
\]
so the scalar integral is ultraviolet finite. Feynman parametrisation
gives
\begin{equation}
I_5
=
-\frac{i\,\Gamma(3)}{16\pi^2m_\phi^6}
F_5^{\rm phys}(r;\hat s_{ij}),
\qquad
r=\frac{M^2}{m_\phi^2},
\label{eq:F5_phys}
\end{equation}
with
\begin{equation}
F_5^{\rm phys}
=
\int d\Pi_5
\left[
r-
\sum_{i<j}x_ix_j
\frac{(K_i-K_j)^2}{m_\phi^2}
-i\epsilon
\right]^{-3}.
\end{equation}
At generic physical kinematics, the function \(F_5^{\rm phys}\)
can develop absorptive parts when physical cuts are crossed. For a
fixed nonexceptional scattering point, its small-\(M\) behavior
gives
\[
\mathcal M_5^{\rm pent}
=
\mathcal O\!\left(
\frac{y^5M^5}{16\pi^2m_\phi^6}
\right),
\]
so the pentagon amplitude vanishes continuously as \(M\to0\).

Similarly, for the pure hexagon,
\[
(N_{\rm FNV},N_{\rm FNP})=(6,0),
\]
the amplitude is
\begin{equation}
i\mathcal M_6^{\rm hex}
=
y^6M^6 I_6,
\qquad
I_6
=
\int\frac{d^dq}{(2\pi)^d}
\frac{1}{D_1D_2D_3D_4D_5D_6}.
\end{equation}
Its superficial degree of divergence is
\[
D_{\rm UV}=4-2(6)=-8,
\]
and, evaluated at the Euclidean symmetric point,
\begin{equation}
I_6
=
\frac{i\,\Gamma(4)}{16\pi^2m_\phi^8}
F_6^E(r),
\qquad
F_6^E(r)
=
\int d\Pi_6\,
\frac{1}{[r+f_6(x_i)]^4},
\label{eq:F6_def}
\end{equation}
where \(f_6(x_i)\) is the positive kinematic polynomial for the
six-point symmetric configuration. The same deterministic numerical
evaluation used for the pure box gives
\[
F_6^E(0)\simeq 0.18.
\]
Near an isolated simplex vertex, the endpoint contribution behaves as
\[
\int_0^1 dt\,
\frac{t^{4}}{(r+t\,g)^{4}}
=
\mathcal O(1)+\mathcal O(r\ln r),
\]
so that
\[
F_6^E(r)=F_6^E(0)+\mathcal O(r\ln r).
\]
No inverse powers of \(r=M^2/m_\phi^2\) compensate the explicit
\(M^6\) FNV mass factor. Hence the pure hexagon amplitude vanishes
as \(M^6/m_\phi^8\) in the symmetry-restoring limit:
\[
\mathcal M_6^{\rm hex}
=
\mathcal O\!\left(
\frac{y^6M^6}{16\pi^2m_\phi^8}
\right).
\]
These contributions are included in the power-counting classification
of Sec.~2.

\subsection{Vacuum contribution and scalar tadpole in the zero-VEV regime}
\label{app:tadpoles}

The one-loop contributions relevant for matching are the vacuum
bubble and the scalar tadpole. The former gives the field-independent
vacuum energy
\begin{equation}
\Delta V_{\rm vac}(M)
=
-\frac{M^4}{32\pi^2}
\left(
\ln\frac{M^2}{\mu^2}-\frac32
\right),
\label{eq:vacuum_bubble_final}
\end{equation}
up to the overall convention for Majorana degrees of freedom. The
scalar tadpole arises from one Yukawa vertex and one FNV contraction,
\begin{equation}
T_\phi
\propto
yM
\int\!\frac{d^dq}{(2\pi)^d}
\frac{1}{q^2-M^2+i\epsilon},
\label{eq:scalar_tadpole_amp}
\end{equation}
which, after dimensional regularisation and \(\overline{\rm MS}\)
subtraction~\cite{tHooft1972dimreg,Bollini1972dimreg}, gives
\[
T_\phi \sim
\frac{yM^3}{16\pi^2}
\left(
1-\ln\frac{M^2}{\mu^2}
\right),
\]
and induces the linear term
\(\Delta V_{\rm tad}=T_\phi\,\phi+T_\phi^*\,\phi^*\).
Perturbation is defined around \(\langle\phi\rangle=0\) by
imposing the tadpole renormalisation condition
\begin{equation}
\Gamma_\phi^{(1),{\rm ren}}(0)=0,
\label{eq:one_point_ren}
\end{equation}
equivalently \(T_\phi+\delta t=0\). This removes the linear term at
the chosen expansion point and fixes the renormalized vacuum around
\(\phi=0\)~\cite{Masina:2025}, without setting \(M=0\) and without
affecting the higher-point \(U(1)\)-violating operators generated
by the fermion loop. The configuration is stable provided the
renormalized curvature at the origin is positive,
\begin{equation}
m_{\phi,\rm eff}^2=
\left.
\frac{\partial^2 V_{\rm eff}^{\rm ren}}
{\partial\phi\,\partial\phi^\dagger}
\right|_{\phi=0}>0.
\label{eq:stability_condition}
\end{equation}
After tadpole renormalisation the lowest non-derivative
symmetry-breaking operator is \(\phi^2+\phi^{*2}\) with
\(\lambda_2\sim y^2M^2\); at cubic order the mixed triangle \((1,2)\)
scales as \(M\) while the pure triangle \((3,0)\) scales as \(M^3\).
The tadpole is therefore removed from the surviving
\((N_{\rm FNV},N_{\rm FNP})\) classification of physical local
vertices, and all explicit-breaking non-derivative operators vanish
as \(M\to0\), in accordance with 't~Hooft
naturalness~\cite{tHooft:1980}.

\section{Coleman--Weinberg potential on the compact Goldstone domain}
\label{app:CW_compact}

The one-loop fermionic Coleman--Weinberg contribution for a single
Majorana eigenstate with field-dependent mass
$
m^2(\theta)
=
m_0^2+M^2+2m_0M\cos\theta
$
is
\begin{equation}
V_{\rm CW}(\theta)
=
-\frac{m^4(\theta)}{32\pi^2}
\left[
\ln\frac{m^2(\theta)}{\mu^2}
-\frac32
\right],
\label{eq:CW_potential_appendix}
\end{equation}
with the Majoron phase on the compact domain
\(0\le\theta<2\pi\), \(\theta\sim\theta+2\pi\).
The first derivative is
\begin{equation}
V'_{\rm CW}(\theta)
=
\frac{m_0M}{8\pi^2}\,
m^2(\theta)
\left[
\ln\frac{m^2(\theta)}{\mu^2}-1
\right]
\sin\theta ,
\label{eq:CW_first_derivative}
\end{equation}
so \(V'_{\rm CW}=0\) requires either \(\sin\theta=0\) or
\(m^2(\theta)=e\,\mu^2\). On the latter branch
\begin{equation}
V''_{\rm CW}(\theta)
=
-\frac{m_0^2M^2}{4\pi^2}\sin^2\theta
<
0 ,
\label{eq:interior_curvature}
\end{equation}
so such configurations are local maxima. Any minimum of the
isolated CW contribution therefore lies at \(\theta=0\) or
\(\theta=\pi\), which correspond to
\begin{equation}
\sigma
\equiv
\cos\theta_s=\pm1 ,
\qquad
m_s^2
\equiv
m^2(\theta_s)
=
(m_0+\sigma M)^2 .
\label{eq:CW_stationary_def}
\end{equation}
At either configuration, \(\sin\theta_s=0\), so the odd derivatives
of \(m^2(\theta)\) vanish. Evaluating the second and fourth
derivatives of \(V_{\rm CW}\) at \(\theta_s\) gives
\begin{equation}
V_2
\equiv
V''_{\rm CW}(\theta_s)
=
\frac{\sigma m_0M\,m_s^2}{8\pi^2}
\left[
\ln\frac{m_s^2}{\mu^2}-1
\right],
\label{eq:CW_V2}
\end{equation}
\begin{equation}
V_3
\equiv
V'''_{\rm CW}(\theta_s)
=
0 ,
\label{eq:CW_V3}
\end{equation}
\begin{equation}
V_4
\equiv
V''''_{\rm CW}(\theta_s)
=
-V_2
-
\frac{3m_0^2M^2}{4\pi^2}
\ln\frac{m_s^2}{\mu^2}.
\label{eq:CW_V4}
\end{equation}
Equations~\eqref{eq:CW_V2} and~\eqref{eq:CW_V4} are exact at either
\(\theta_s=0\) or \(\theta_s=\pi\), at arbitrary renormalization
scale \(\mu\). Because \(m^2(\theta_s+\delta\theta)=m^2(\theta_s-\delta\theta)\),
all odd derivatives of \(V_{\rm CW}\) vanish at \(\theta_s\) exactly,
to all orders in \(M\), and the Taylor expansion takes the form
\begin{equation}
V_{\rm CW}(\theta_s+\delta\theta)
=
V_0
+
\frac12 V_2\,\delta\theta^2
+
\frac{1}{24}V_4\,\delta\theta^4
+
\mathcal O(\delta\theta^6).
\label{eq:CW_local_expansion}
\end{equation}

With the canonically normalized fluctuation \(a=v\,\delta\theta\),
this may be written as
\(V_{\rm CW}=V_0+\frac12\kappa_{a,\rm CW}a^2
+\frac{1}{4!}g_{4,\rm CW}a^4+\cdots\), where the curvature at the
stationary configuration is
\begin{equation}
\kappa_{a,\rm CW}
\equiv
\frac{V_2}{v^2}
=
\frac{\sigma m_0M\,m_s^2}{8\pi^2 v^2}
\left[
\ln\frac{m_s^2}{\mu^2}-1
\right],
\label{eq:CW_curvature}
\end{equation}
and the quartic coupling is
\begin{equation}
g_{4,\rm CW}
=
-\frac{\kappa_{a,\rm CW}}{v^2}
-
\frac{3m_0^2M^2}{4\pi^2 v^4}
\ln\frac{m_s^2}{\mu^2}.
\label{eq:CW_quartic}
\end{equation}
If the chosen stationary configuration is a local minimum, then
\(\kappa_{a,\rm CW}>0\) and may be identified with the curvature
mass squared of the isolated CW contribution,
\(m_{a,\rm CW}^2=\kappa_{a,\rm CW}\). At a maximum,
\(\kappa_{a,\rm CW}<0\), and the identification does not apply.

For \(M/m_0\ll1\), using \(m_s^2=m_0^2+\mathcal O(M)\) and
\(\ln(m_s^2/\mu^2)=\ln(m_0^2/\mu^2)+\mathcal O(M/m_0)\),
\begin{equation}
\kappa_{a,\rm CW}
=
\frac{\sigma m_0^3M}{8\pi^2 v^2}
\left[
\ln\frac{m_0^2}{\mu^2}-1
\right]
+
\mathcal O\!\left(\frac{m_0^2M^2}{v^2}\right),
\label{eq:CW_curvature_leading}
\end{equation}
and \(g_{4,\rm CW}=-\kappa_{a,\rm CW}/v^2
+\mathcal O(m_0^2M^2/v^4)\).
The local expansion assumes \(m_s^2>0\). At the exceptional point
\(m_0=M\), \(\theta_s=\pi\), the fermion becomes massless and the
mass eigenvalue behaves as \(m^2\approx M^2\delta\theta^2\) locally,
so that
\(V_{\rm CW}\sim-(m_0^4/32\pi^2)\delta\theta^4\ln\delta\theta^2\).
The CW potential itself remains finite, but the ordinary
fourth-order Taylor expansion is nonanalytic and requires separate
treatment. This configuration lies outside the regime
\(M/m_0\ll1\) considered in the main text.

\end{document}